\documentclass[prx,twocolumn,showpacs,amsmath,amssymb,superscriptaddress]{revtex4-2}

\usepackage{amsmath,amssymb,amsfonts,bm,color,graphicx,tabularx}

\usepackage[unicode=true,colorlinks=true]{hyperref}

\usepackage[etex=true,export]{adjustbox}
\usepackage{makecell}
\usepackage{multirow}
\usepackage{amsmath}

\hypersetup{linkcolor=blue, citecolor=blue,urlcolor=blue}
\usepackage{tabularx,multirow,array,diagbox}
\usepackage{adjustbox}

\newcommand{\beq}{\begin{equation}}
\newcommand{\eeq}{\end{equation}}
\newcommand{\beql}{\begin{equation*}}
\newcommand{\eeql}{\end{equation*}}
\newcommand{\beqn}{\begin{eqnarray}}
\newcommand{\eeqn}{\end{eqnarray}}

\begin{document}
\title{ Tunable quantum metric and band topology in bilayer Dirac models}

\author{Xun-Jiang Luo}
\affiliation{Department of Physics, Hong Kong University of Science and Technology, Clear Water Bay, 999077 Hong Kong, China}

\author{Xing-Lei Ma}
\affiliation{Department of Physics, Hong Kong University of Science and Technology, Clear Water Bay, 999077 Hong Kong, China}

\author{K. T. Law}
\email{phlaw@ust.hk}
\affiliation{Department of Physics, Hong Kong University of Science and Technology, Clear Water Bay, 999077 Hong Kong, China}

\begin{abstract}

The quantum metric, a fundamental component of quantum geometry, has attracted broad interest in recent years due to its critical role in various quantum phenomena. Meanwhile, band topology, which serves as an important framework in condensed matter physics, has led to the discovery of various topological phases. In this work, we introduce a “bilayer”  Dirac model that allows precise tuning of both properties.
 Our approach combines two Dirac Hamiltonians with distinct energy scales; one producing relatively dispersive bands and the other yielding relatively flat bands. The dispersive and flat bands are weakly coupled by hybridization $\lambda$. By inducing a band inversion in the “layer” subspace, we achieve flexible tuning of band topology across all Altland-Zirnbauer symmetry classes and quantum metric scaling as $g\propto 1/\lambda^2$ near the band inversion point. Using the “bilayer” Su-Schrieffer-Heeger  model, we investigate the localization properties of gapless boundary states, which are affected by quantum metric. Our work lays a foundation for exploring the interplay between band topology and quantum metric.

\end{abstract}
\maketitle

\section{Introduction}

The quantum geometric tensor \cite{2010arXiv1012.1337C,10.1093/nsr/nwae334,2025arXiv250100098Y}, which comprises the Berry curvature and quantum metric, serves as a fundamental framework to characterize the geometric properties of Bloch states in condensed matter systems.  Berry curvature, the imaginary part of this tensor, governs topological phenomena such as the quantized anomalous Hall effect \cite{RevModPhys.82.1959}. This effect arises when the integral of Berry curvature over the Brillouin zone yields a nonzero Chern number, which protects gapless boundary states against perturbations \cite{RevModPhys.82.3045,Qi2011}. This intrinsic relationship facilitates the development of topological band theory \cite{Chiu2016}, a cornerstone of modern condensed matter physics, and has led to the discovery of various topological phases, including topological insulators \cite{Kane2005,Bernevig2006,PhysRevLett.98.106803}, topological superconductors  \cite{Fu2008,Lutchyn2010,Luo20241}, and higher-order topological phases \cite{Benalcazar2017a,Benalcazar2017,Song2017,Luo2021a,luo2024,Luo2023a}. In contrast,  quantum metric, the real part of the quantum geometric tensor, quantifies the distance between quantum states in momentum space. It plays a pivotal role in a wide range of physical phenomena beyond the scope of Berry curvature, such as the nonlinear Hall effect \cite{gaoyang2014,LiuHuiying2021,wangchong2021,anyuanguo2023}, optical absorption spectra \cite{SouzaIvo2000,GaoYang2019,Jankowski2024,Onishi2024,Komissarov2024,Verma2024step,Verma2024a,Resta2025}, entanglement entropy \cite{PaulNisarga2024,TamPokMan2024,ZhouLongwen2024,Kruchkov2024,wuxiaochuan2024}, and flat band superconductivity \cite{Peotta2015,Julku2017,XieFang2020,HerzogArbeitman2022,HofmannJohannesS2023,Peotta2023,2024sunziting}. Recent research has revealed a profound interplay between band topology and quantum metric \cite{RoyRahul2014,OzawaTomoki2021,OkumaNobuyuki2024,KwonSoonhyun2024,DingHaiTao2024,Romeral2024,Yu2025,chiupokman2025,Jankowski2025}. For example, topological invariants such as the Chern number \cite{RoyRahul2014}, Euler number \cite{KwonSoonhyun2024,Jankowski2025a},  Wilson loop invariant \cite{Yu2025}, and spin Chern number \cite{Jankowski2025} impose a lower bound on  the quantum metric. This connection suggests that the quantum metric and band topology are not merely parallel concepts but are intricately intertwined, collectively shaping the electronic properties of materials.

\begin{figure}[t]
\centering
\includegraphics[width=3.3in]{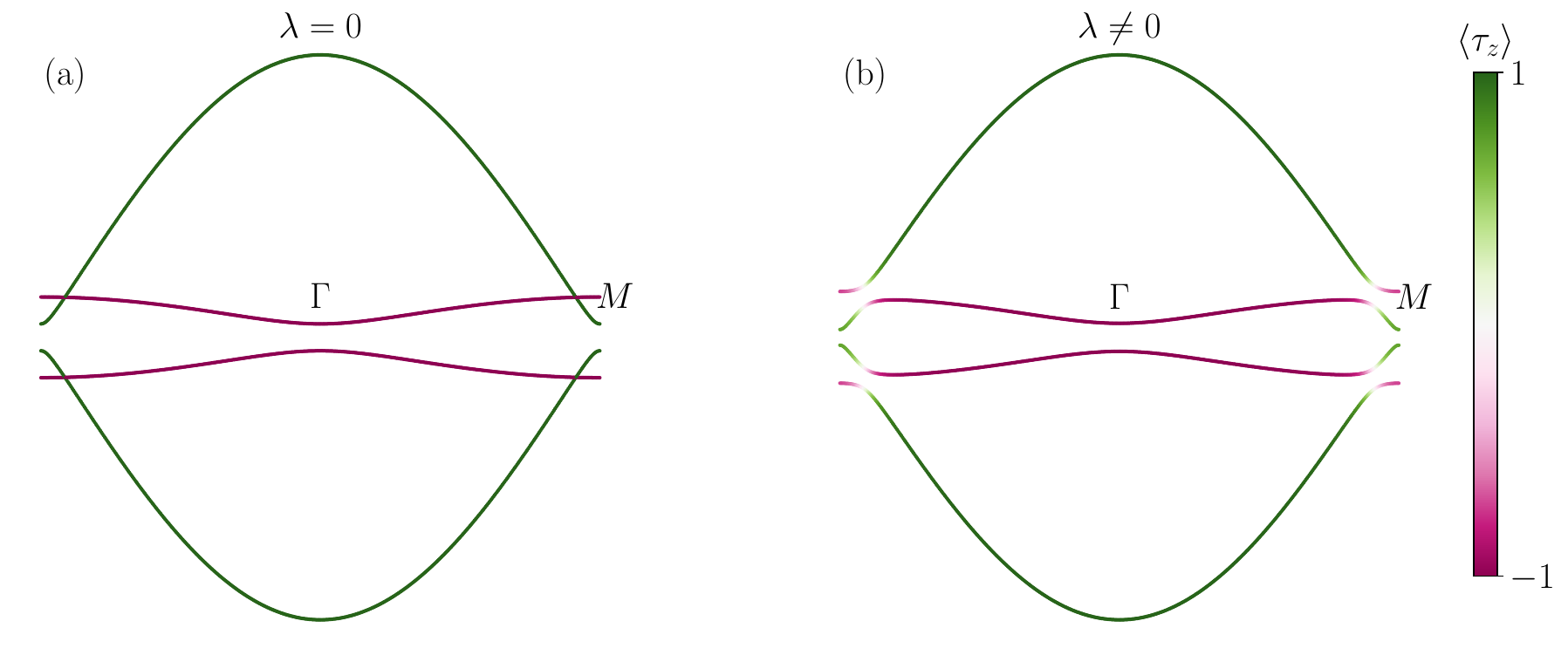}
\caption{(a) The relatively dispersive (green) and flat (red) bands. (b) The illustration of band inversion between the dispersive and flat bands at the $M$ point in the “layer” subspace. The colorbar encodes the expectation value of the “layer” pseudospin $\tau_z$. For the plot, the common model parameters are taken as $\epsilon_1=2$, $\epsilon_2=-0.4$, $m_1=0.2$, $m_2=0.6$ for the “bilayer” SSH model. In (b), $\lambda=0.2$ is taken.  }
\label{fig1}
\end{figure}

Systems with nearly flat bands provide an ideal platform for investigating quantum phenomena governed by the quantum metric \cite{Mitscherling2022,Yuhang2024,Xuzhe2024arXiv,Xuzhe2024,ChunWang2024,Kevin2025}, as the suppression of kinetic energy enhances its significance. For instance, the quantum metric can induce superfluid weight, which is dominant in flat band systems. Moreover, recent studies \cite{ChenShuaiA2024,lizhong2024,Hu2025,2025arXivxinglei}  have demonstrated that the quantum metric provides an alternative length scale that critically influences the physical properties of flat band systems. In particular, a more recent study \cite{xingyao2024}, employing the Lieb-Kitaev model featuring isolated and topological flat bands, showed that the quantum metric establishes a new localization length for topological boundary states in addition to the conventional localization length governed by Fermi velocity.

\begin{figure}[t]
\centering
\includegraphics[width=3.4in]{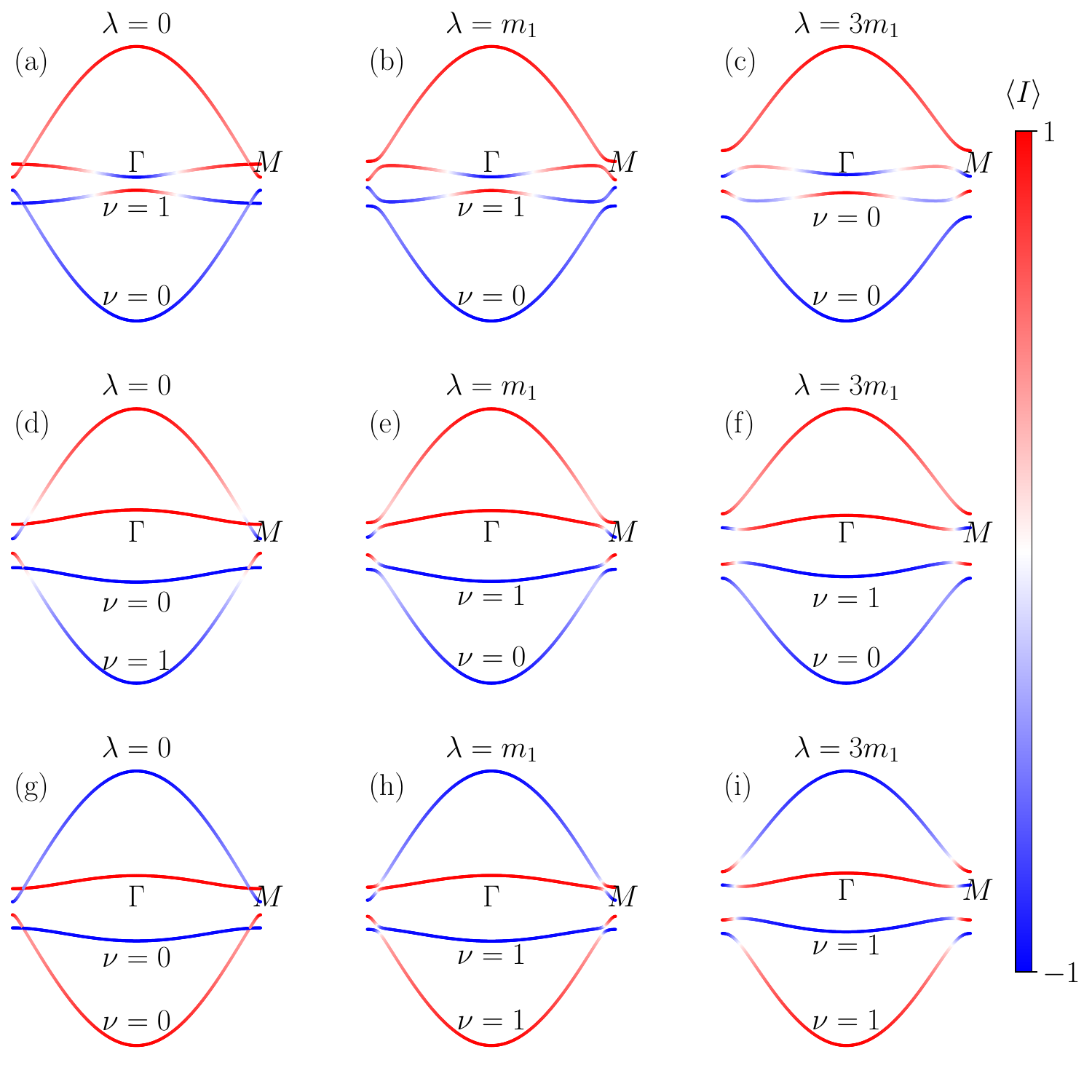}
\caption{Three distinct cases for realizing topological flat bands. The colorbar encodes the expectation value of inversion operator $I$.  The band topology are characterized by the topological invariant $\nu$.
 The same model parameters are used for (a)-(c) as those used in Fig.~\ref{fig1}.   In (d)-(f), we take $\epsilon_1=2$, $\epsilon_2=0.2$, $m_1=-0.2$, and $m_2=0.6$ for the plot.  In (c)-(g), we take  $\epsilon_1=-2$, $\epsilon_2=0.2$,  $m_1=-0.2$, and $m_2=0.6$ for the plot.
 }
\label{fig2}
\end{figure}

Dirac models are essential for describing topological phases \cite{Ryu2010}.
In this work, we introduce a “bilayer” Dirac model to construct topological flat bands with tunable quantum metric across the full Altland-Zirnbauer (AZ) tenfold symmetry classes \cite{Chiu2016}. 
Our approach utilizes two Dirac Hamiltonians with distinct energy scales: one generates relatively dispersive bands, while the other produces nearly flat bands. By tuning the model parameters, we ensure that the dispersive bands intersect the nearly flat bands near a time-reversal-invariant point. A small “interlayer” coupling $\lambda$ then induces a band inversion in the ``layer'' subspace, as illustrated in Fig.~\ref{fig1}. This mechanism allows for precise control over the band topology, characterized by Fu-Kane parity invariants \cite{FuLiang2027inversion}, as shown in Fig.~\ref{fig2}, while simultaneously enabling the quantum metric to be tuned, scaling as $\propto 1/\lambda^2$, as demonstrated in Fig.~\ref{fig3}. This tunability leads to a significant quantum weight, the momentum-space integral of the quantum metric, particularly for small values of $\lambda$. Using the constructed “bilayer” Su-Schrieffer-Heeger (SSH) model as an example, we show that the quantum metric can induce long-range decay behavior for the gapless boundary states.

This paper is organized as follows. In Sec.~\ref{II},  we present the construction method of “bilayer” Dirac models.  In Sec.~\ref{III}, we show the tunable band topology in “bilayer” Dirac models. In Sec.~\ref{V}, we demonstrate the tunable quantum metric in “bilayer” Dirac models. In Sec.~\ref{IV}, we study the localized properties of end states of the “bilayer” SSH model. 
In Sec.~\ref{VI}, we present a brief discussion and summary.
Appendix \ref{Appendix A} complements the main text.

\section{Construction method}
\label{II}
Before presenting the construction of “bilayer” Dirac models, we first provide a brief overview of Dirac model and present its topological characterization by the generalized Fu-Kane parity invariants \cite{Fu200C,FuLiang2027inversion} .

\subsection{Dirac model}

For a $d$-dimensional ($d$D) Dirac model, the Hamiltonian is expressed as \cite{2024BTIluo}
\begin{equation}
H(\bm k) = \sum_{i=1}^{d} \sin k_ia \, \gamma_{i}^{(g)} + M(\bm k) \, \gamma_{d+1}^{(g)},
\label{dirac}
\end{equation}
where $M(\bm k) = m + \sum_{i=1}^{d} (1 - \cos k_ia)$, with $m$ being a model parameter and $a$ being the lattice constant. The $2^g \times 2^g$ Gamma matrices $\gamma_j^{(g)}$ satisfy the anti-commutation relation $\{\gamma_j^{(g)}, \gamma_{j'}^{(g)}\} = 2 \delta_{jj'}{ I}^{(g)}$ for $j, j' = 1, \dots, d+1$, where ${ I}^{(g)}$ is a $2^g \times 2^g$  identity matrix, and $d \leq 2g$.  When present, the time-reversal ($T$), particle-hole ($P$), and chiral ($C$) symmetries act on $H(\bm k)$ as:
\begin{align}
T H(\bm k) T^{-1} &= H(-\bm k), \nonumber \\
P H(\bm k) P^{-1} &= -H(-\bm k), \nonumber \\
C H(\bm k) C^{-1} &= -H(\bm k).
\end{align}
The presence or absence of these three local symmetries categorizes gapped systems into AZ tenfold symmetry classes \cite{Chiu2016}.

For a given symmetry class and dimension $d$, $H$ describes a stronger TI or TSC if two conditions are met: (i) $H$ lacks a symmetry-preserving
extra mass term; (ii) the node surface of $M(\bm k)$ (where $M = 0$) wraps an odd number of time-reversal invariant points. These conditions ensure that the gapped phase cannot be adiabatically connected to a trivial atomic limit phase, where the mass term reduces to a constant $M_0(\bm k) = \mathcal{M}$, without closing the bulk energy gap or breaking certain symmetry. Consequently, the nontrivial band topology is intrinsically related to the gap closing process of $H$. Moreover, $H(\bm k)$ only closes its energy gap at time-reversal invariant points and $H$ respects the inversion symmetry ${I} = \gamma_{d+1}$, where ${I} H(\bm k) {I}^{-1} = H(-\bm k)$. Therefore, the topology of the occupied bands can be characterized by the generalized Fu-Kane invariant \cite{Fu200C,FuLiang2027inversion}:
\begin{equation}
(-1)^\nu = \prod_{i=1}^{2^d} \delta_i,
\label{tv}
\end{equation}
where $\delta_i=\pm 1$ denotes the inversion eigenvalue at the $i$-th time-reversal invariant point, and $\nu$ distinguishes trivial ($\nu = 0$) and nontrivial ($\nu = 1$) phases.


The realization of topological insulators and superconductors by $H$ across the full AZ tenfold symmetry classes can be exemplified by the following paradigmatic models. For $d = 1$ and $g = 1$, $H$ reduces to the two-band SSH model \cite{Su1979} in the chiral-symmetric AIII class, which hosts robust end states. For $d = 2$ and $g = 1$, $H$ corresponds to the Qi-Wu-Zhang model \cite{Qi2006}, realizing a Chern insulator in the A symmetry class. For $d = 2$ and $g = 2$, $H$ represents the Bernevig-Hughes-Zhang model \cite{Bernevig2006}, describing a time-reversal invariant topological insulator in the AII symmetry class.

\begin{figure}[t]
\centering
\includegraphics[width=3.4in]{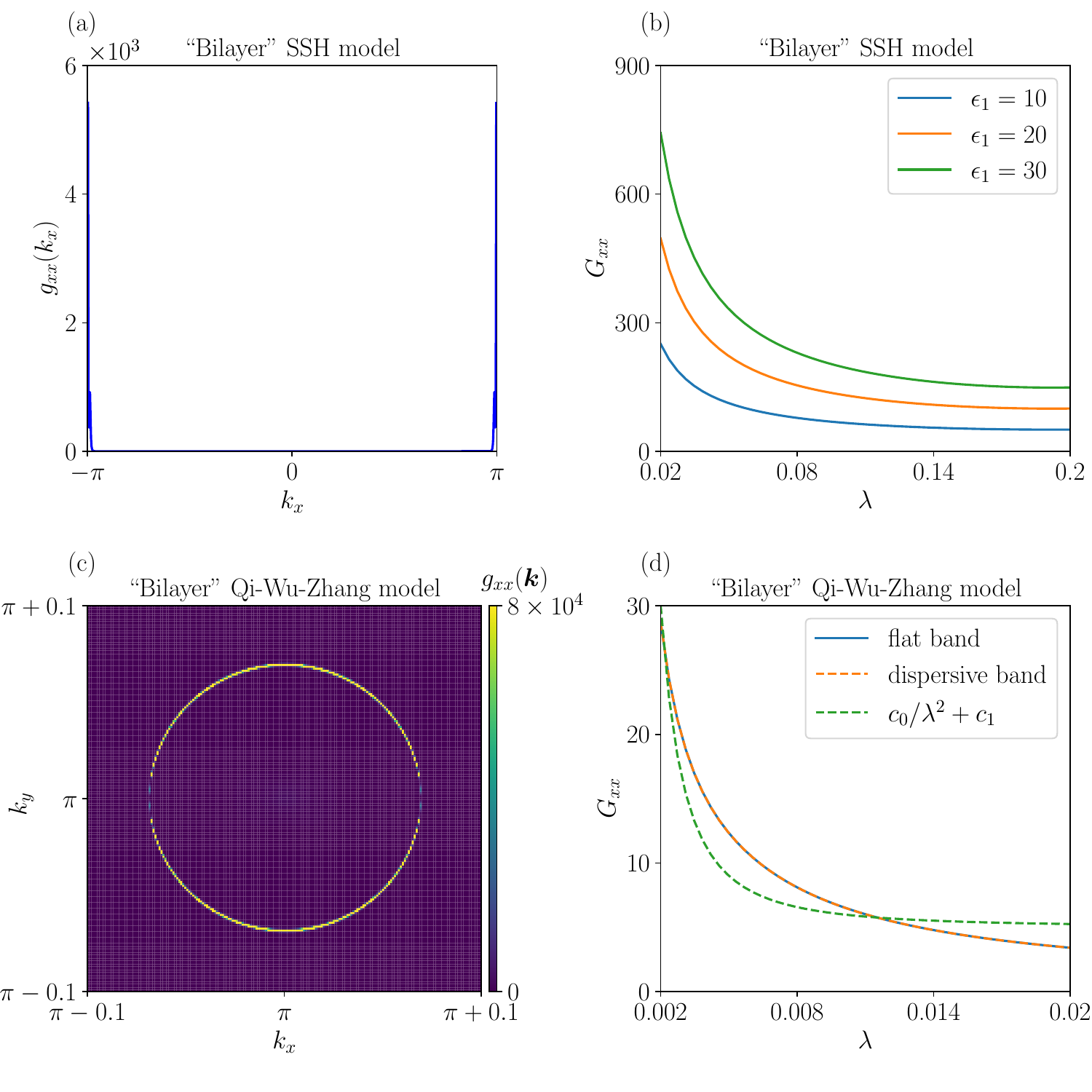}
\caption{(a) The distribution of the quantum metric $g_{xx}(k_x)$ for the “bilayer” SSH model. (b) The quantum weight as a function of $\lambda$ under different settings of $\epsilon_1$.  (c) The distribution of the quantum metric $g_{xx}(k_x,k_y)$ for the “bilayer” Qi-Wu-Zhang model. (d) The quantum weight of the dispersive (flat) bands and their fitting results for the “bilayer” Qi-Wu-Zhang model as a function of $\lambda$, with $c_0=0.0001$ and $c_1=5$. The common model parameters are taken as $\epsilon_2=-0.4$, $m_2=0.6$, and $m_1=0.2$. In (a), we take $\epsilon_1=20$ and $\lambda=0.2$. In (c), we take $\epsilon_1=20$ and $\lambda=0.02$. In (d), we take $\epsilon_1=20$.
 }
\label{fig3}
\end{figure}

\subsection{“Bilayer”  Dirac models}
\label{BDM}
To generally construct topological flat bands with tunable quantum metric, we introduce a “bilayer” Dirac model. The model Hamiltonian is given by
\begin{equation}
\mathcal{H} = \begin{pmatrix}
\epsilon_1 H_{1} & \lambda I^{(g)} \\
\lambda I^{(g)} & \epsilon_2 H_{2}
\end{pmatrix},
\label{dirac2}
\end{equation}
where $H_1$ and $H_2$ are standard Dirac Hamiltonians with the same form as $H$ in Eq.~\eqref{dirac}. In $H_1$ and $H_2$, identical representations of the Gamma matrices are chosen and therefore they have the identical symmetries, including $T$, $P$, and $C$, if present, and inversion symmetry $I$. The Dirac masses of $H_1$ and $H_2$ are, respectively, chosen as
\begin{align}
M_1(\bm k) &= \frac{m_1}{\epsilon_1} + \sum_{i=1}^{d} (1+ \cos k_ia), \nonumber \\
M_2(\bm k) &= \frac{m_2}{\epsilon_2} + \sum_{i=1}^{d} (1 + \cos k_ia).
\end{align}
Here, $\epsilon_{1,2}$ and $m_{1,2}$ are model parameters. $\epsilon_1$ and $\epsilon_2$ set the energy scales of $H_1$ and $H_2$, respectively. Both $H_1$ and $H_2$ have two nondegenerate ($g=1$) or degenerate ($g>1$) Dirac bands, and the ratio between the energy band bandwidths of $H_1$ and $H_2$ is $\epsilon_1 / \epsilon_2$. We set $\epsilon_1 \gg \epsilon_2$, such that the two Dirac bands of $H_1$ and $H_2$ are relatively dispersive and flat, respectively. To ensure that the dispersive band crosses the flat band near the point $M = (\pi, \dots, \pi)$, we require $|m_1| < |m_2|$. When a small coupling $\lambda$ is introduced between the dispersive and flat bands, a band inversion is created in the ``layer'' subspace, as illustrated in Fig.~\ref{fig1}. This band inversion is crucial for tuning the band topology and quantum metric of $\mathcal{H}$, as we will demonstrate in the following.

\section{Tunable band topology}
\label{III}
In the study of band topology, identifying the symmetry class of Hamiltonian is crucial, as it determines its topological classification \cite{Chiu2016}. To analyze the symmetry class, we rewrite $\mathcal{H}$ as
\beqn
\mathcal{H}=(\tau_0+\tau_z)/2H_{1}+(\tau_0-\tau_z)/2H_{2}+\delta\tau_x,
\eeqn
where $\tau_{0,x,y,z}$ are the Pauli matrices acting on the “layer” subspace which are expanded by the eigenbasis of $H_{1}$ and $H_2$. With this form, it can be readily shown that there is an exact one-to-one correspondence between the three local symmetries for $\mathcal{H}$ and $H_{1}$,
\beqn
\mathcal{T}=\tau_0\otimes T,\quad
\mathcal{P}=\tau_z\otimes P, \quad
\mathcal{C}=\tau_z\otimes C.
\eeqn
Here, $\mathcal{T}$, $\mathcal{P}$, and $\mathcal{C}$ denote the time-reversal, particle-hole, and chiral symmetries of $\mathcal{H}$. Therefore, the Hamiltonians $\mathcal{H}$ and $H_{1,2}$ belong to the identical symmetry class. 
In addition to the $T$, $C$, $P$ symmetries, there is also a correspondence between the inversion symmetry  $\mathcal{I}=\tau_0\otimes I$, where $\mathcal{I}\mathcal{H}(\bm k)\mathcal{I}^{-1}=\mathcal{H}(-\bm k)$. Therefore,  inversion symmetries $\mathcal{I}$ and $I$ have identical eigenvalues at time-reversal invariant momenta and the band topology of $\mathcal{H}$ can still be characterized by the topological invariant $\nu$ defined by Eq.~\eqref{tv}.

The band topology of $H_1$ and $H_2$ is determined by the model parameters $m_1$ and $m_2$, respectively, and characterized by the invariant $\nu$. For example, when $-2d < m_i / \epsilon_i < -2d + 2$ with $i = 1, 2$, $H_i$ exhibits a band inversion at the $\Gamma = (0, \dots, 0)$ point, resulting in topological bands characterized by $\nu = 1$. The band topology can be further tuned by the band inversion ($\lambda\neq 0$) between the dispersive and flat bands.  Generally, there are three distinct ways to realize topological flat bands for $\mathcal{H}$:  
(i) Input topological flat bands from $H_2$, and the band inversion between the dispersive and flat bands does not change the band topology, as illustrated in Figs.~\ref{fig2}(a) and \ref{fig2}(b);  
(ii) $H_1$ and $H_2$ is topologically nontrivial and trivial, and the band inversion at the $M$ point change the band topology, enabling the realization of topological flat bands and trivial dispersive bands, respectively, as illustrated in Figs.~\ref{fig2}(d) and \ref{fig2}(e); 
(iii) Both $H_1$ and $H_2$ are topologically trivial, and the band inversion at the $M$ point enables the realization of topological dispersive and flat bands, respectively, as illustrated in Figs.~\ref{fig2}(g) and \ref{fig2}(h).
Case (i) can be realized by setting $m_1 / \epsilon_1 > 0$, $-2d < m_2 / \epsilon_2 < -2d + 2$, and $m_{1}m_{2}>0$. Case (ii) can be realized by setting $-2 < m_1 / \epsilon_1 < 0$, $m_2 / \epsilon_2 >0$, and $m_{1}m_{2}<0$. Case (iii) can be realized by setting $m_{1,2} / \epsilon_{1,2} > 0$ and $m_{1}m_{2}<0$. For these three cases, it can be shown that the obtained flat bands under small $\lambda$ are topologically nontrivial and characterized by $\nu = 1$. Therefore, $\mathcal{H}$ can realize topological flat bands across the full AZ symmetry classes by following the construction procedure and choosing different dimensions and symmetry classes for $\mathcal{H}$. Here, we assume that $\lambda$ is small compared to the energy gap of $H_1$ at the $M$ point, namely $m_1$. 
When $\lambda > m_1$, the band topology for case (i) could be further changed by closing the energy gap at the $M$ point, as shown in Fig.~\ref{fig2}(c). While for cases (ii) and (iii), we find that the band topology is relatively robust against the variation of $\lambda$, as shown in Figs.~\ref{fig2}(f) and \ref{fig2}(i).  Notably, although the band inversion between the dispersive and flat bands is not necessary to realize nontrivial band topology, it is essential for tuning the quantum metric, as we will show in the following.


\begin{figure}[t]
\centering
\includegraphics[width=3.4in]{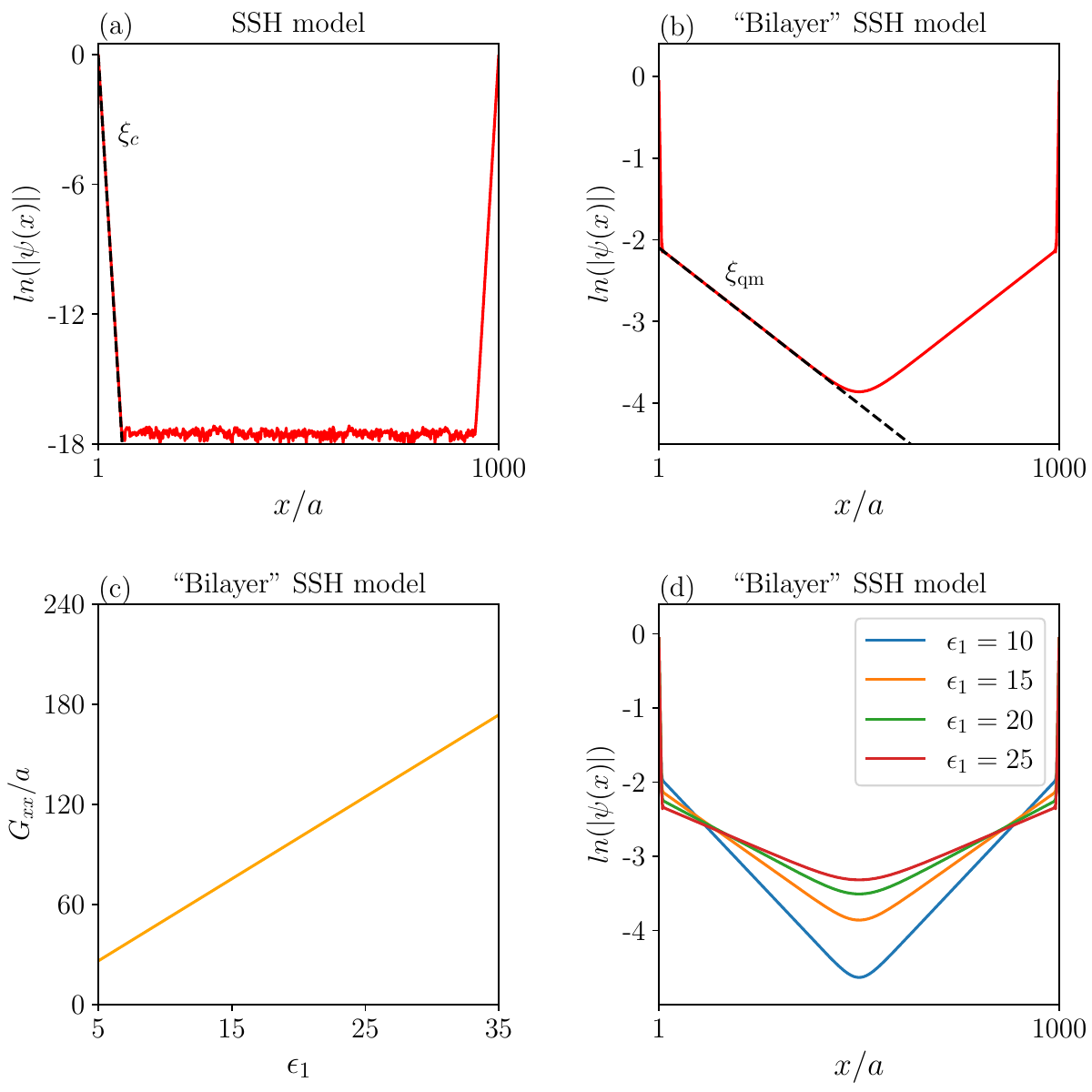}
\caption{(a) The plot of $\text{ln}|\psi(x)|$ for end states in the two-band SSH model. (b) The plot of $\text{ln}|\psi(x)|$ for end states in the “bilayer” SSH model. (c) The quantum weight $G_{xx}$ as a function of $\epsilon_1$ for the “bilayer” SSH model. (d) The plot of $\text{ln}|\psi(x)|$ for end states in the “bilayer” SSH model under different setting of $\epsilon_1$. The common model parameters are taken as $m_2=0.6$, $\epsilon_2=-0.4,\lambda=m_1=0.2$. In (b), we take $\epsilon_1=15$.
 }
\label{fig4}
\end{figure}

\section{Tunable quantum metric}
\label{IV}
Within the framework of Dirac models across arbitrary dimensions, it has been demonstrated that nontrivial band topology inherently establishes a lower bound for the quantum metric \cite{Mera2022}. However, this lower bound, determined by topological invariants, is typically small. To effectively investigate the quantum metric effect and its interplay with band topology, it is essential to construct topological flat bands with both tunable and significantly enhanced quantum metric values.  We now demonstrate the tunable quantum metric in $\mathcal{H}$.

When $g > 1$, the energy bands of $\mathcal{H}$ exhibit degeneracy, necessitating the inclusion of the non-Abelian quantum geometric tensor. This tensor is defined as \cite{2010arXiv1012.1337C,Kwon2024}
\begin{equation}
Q_{\mu \nu}^{ij}(\mathbf{k}) = \left\langle \partial_\mu u_{n_i}(\mathbf{k}) \Big| (1 - P(\mathbf{k})) \Big| \partial_\nu u_{n_j}(\mathbf{k}) \right\rangle,
\end{equation}
where $\mu, \nu = x, y, z, \dots$ represent the spatial coordinates, and $\partial_\mu = \frac{\partial}{\partial k_\mu}$ denotes the derivative with respect to the momentum components. The indices $i$ and $j$ label the possible degeneracy of the $n$-th band. The projection operator $P(\mathbf{k})$ is given by
\begin{equation}
P(\mathbf{k}) = \sum_{i=1}^{2^{(g-1)}} \left| u_{n_i}(\mathbf{k}) \right\rangle \left\langle u_{n_i}(\mathbf{k}) \right|,
\end{equation}
where \(\left\{ \left| u_{n_1, \dots, n_2^{(g-1)}} (\mathbf{k}) \right\rangle \right\}\) denote the degenerate eigenstates of the $n$-th band of $\mathcal{H}$. The real part of $Q_{\mu \nu}$ is the non-Abelian quantum metric, and its trace yields the standard quantum metric:
\begin{equation}
g_{\mu \nu}(\mathbf{k}) \equiv \frac{1}{2} \operatorname{Tr} \left[ Q_{\mu \nu} + Q_{\nu \mu} \right],
\end{equation}
where the trace \(\operatorname{Tr}\) is taken over the degenerate band indices. The quantum metric $g_{\mu \nu}(\mathbf{k})$ is invariant under gauge transformations for complex wave functions of degenerate bands. For numerical convenience, we use the equivalent expression \cite{2010arXiv1012.1337C}
\begin{equation}
g_{\mu \nu} = \sum_{m \neq n, i,j} \frac{\left\langle\phi_{n_i}\right| \partial_\mu H\left|\phi_{m_j}\right\rangle\left\langle\phi_{m j}\right| \partial_\nu H\left|\phi_{n_i}\right\rangle}{\left(E_n - E_m\right)^2}.
\label{qmc}
\end{equation}
This formulation provides an efficient approach for numerically calculating the quantum metric.

In $\mathcal{H}$,  the quantum metric $g_{\mu\nu}$ can be flexibly tuned by the coupling parameter $\lambda$. Near the band touch point between the dispersive and flat bands [Fig.~\ref{fig1}(a)], denoted by  $\bm k_0$, there is a stronger mixing of eigenstates in the “layer” subspace. Therefore, eigenstates have a larger sensitivity to momentum variations near $\bm k_0$, which results in an increased quantum metric. In particular, when $\lambda \ll \epsilon_2 \ll \epsilon_1$, the energy gap between the dispersive and flat bands near $\bm{k}_0$ is proportional to $\lambda$. Consequently, the quantum metric exhibits a dependence of $1/\lambda^2$  according to Eq.~\eqref{qmc}. For a small value of $\lambda$, quantum metric displays a sharp peak around  $\bm{k}_0$, significantly contributing to its integral, defined as the quantum weight \cite{Onishi2024a,Onishi2024b},
\beqn
G_{\mu\nu} = 2\pi \int \frac{\mathrm{d}^d k}{(2\pi)^d} g_{\mu\nu}(\boldsymbol{k}),
\eeqn
which has dimension $a^{2-d}$.
Thus, we expect $G$ to scale approximately as $1/\lambda^2$ under condition $\lambda \ll \epsilon_2 \ll \epsilon_1$.

We utilize the topological flat bands realized in case (i) [Fig.~\ref{fig2}(b)] to investigate the tunable quantum metric in $\mathcal{H}$. In Fig.~\ref{fig3}(a), we plot the distribution of the quantum metric $g_{xx}(k_x)$ in momentum space, using the two-band SSH model for $H_{1,2}$. The quantum metric $g_{xx}(k_x)$ exhibits a sharp peak near $\bm k_0$, with a maximum value scaling as $1/\lambda^2$. In Fig.~\ref{fig3}(b), we further plot the quantum weight $G_{xx}$ as a function of $\lambda$ under different settings of $\epsilon_1$, while keeping $m_1$, $m_2$, and $\epsilon_2$ fixed. As the bandwidth of the dispersive band ($\epsilon_1$) increases, $G_{xx}$ increases. This behavior can be directly related to the larger variation of $\partial_{\mu} H$ with the increasing of $\epsilon_1$ according to Eq.~\eqref{qmc}. In Fig.~\ref{fig3}(c), we present the distribution of the quantum metric $g_{xx}$ in momentum space, adopting the Qi-Wu-Zhang model \cite{Qi2006} for $H_{1,2}$. In Fig.~\ref{fig3}(d), we plot the quantum weight $G_{xx}$ for the dispersive and flat bands, observing that their values are nearly equal. Moreover, $G_{xx}$ scales approximately to $1/\lambda^2$, consistent with our expectations.

\section{Quantum metric effect on boundary states}
\label{V}
Topological bands guarantee the existence of gapless boundary states, known as the bulk-boundary correspondence \cite{RevModPhys.82.3045,Qi2011}. In the nontrivial parameter region, $H_2$ hosts gapless states located at $(d-1)$-dimensional boundary, which are described by the wavefunction $
\psi(\bm{k}_{\parallel}, r) = A e^{-r/\xi_{\text{c}}(\bm{k}_{\parallel})} | \phi(\bm{k}_{\parallel}) \rangle $.
Here, $\bm{k}_{\parallel}$ denotes the parallel momentum along the periodic boundary directions and $r$ represents the real-space coordinate along the open boundary direction.
 $\xi_{\text{c}}(\bm{k}_{\parallel})$ is the momentum-dependent localization length determined by the model parameters.
$|\phi(\bm{k}_{\parallel})\rangle$ encodes the spinor structure of the boundary states and $A$ is the normalization constant.
At the boundary Dirac point ($\bm k_{||}=0$), $H_2$ reduces to an extended SSH model \cite{Luo2023a}, namely ($g-1$) copies of the two-band SSH model, and it can be shown that $a/\xi_{\text{c}}=\ln \left(|\epsilon_2 / t|\right)$, where $t=m_2+2\epsilon_2(d-1)$ \cite{Asboth2015ASC}. In Fig.~\ref{fig4} (a),  the exponential decay of the edge-state amplitudes across lattice sites for the two-band SSH model is plotted.


For isolated topological bands which are separated from dispersive bulk states,  recent work of Refs.~\onlinecite{xingyao2024,2025arXivxinglei} revealed that the quantum metric can introduce a geometry localization length $\xi_{\text{g}}$ for gapless boundary states. Since $\mathcal{H}$ can realize topological flat bands with tunable quantum metric across all AZ symmetry classes, the constructed models provide a testbed for demonstrating quantum metric effect on boundary states. Generally, $\mathcal{H}$ hosts the boundary states described by
\beqn
\psi(\bm{k}_{\parallel},r)=A_{\text{c}}e^{-r/\xi_{\text{c}}(\bm{k}_{\parallel})}+A_{\text{g}}e^{-r/\xi_{\text{g}}(\bm{k}_{\parallel})},
\eeqn
where $A_{\text{c}}$ and $A_{\text{g}}$ represent the amplitudes of the two distinct components of the wave function. $\xi_{\text{c}}$ is the conventional localization length determined by the 
Fermi-velocity $v_F$ and the energy gap $\Delta$ between the topological conduction and valence bands. In nearly flat bands ($\epsilon_1 \gg \epsilon_2$), $\xi_{\text{c}}$ approaches zero, the long-range decay behavior of gapless boundary states is mainly determined by $\xi_{\text{g}}$. At the boundary Dirac point ($\bm{k}_{\parallel}=0$), $\mathcal{H}$ reduces to a  “bilayer” extended SSH model, which determines the decayed behavior of the gapless boundary states of $\mathcal{H}$ across all the symmetry classes and dimensions.

We now investigate the localization properties of the end states in the “bilayer” SSH model described by $\mathcal{H}$. In contrast to the edge states of $H_2$ shown in Fig.~\ref{fig4}(a), the end states of $\mathcal{H}$ exhibit long-range decay behavior characterized by a large value of $\xi_{g}$, as depicted in Fig.~\ref{fig4}(b). Fig.~\ref{fig4}(c) plots the quantum weight $G_{xx}$ as a function of $\epsilon_1$, which increases linearly with $\epsilon_1$. Correspondingly, the geometry localized length $\xi_{g}$ also increases with $\epsilon_1$, as shown in Fig.~\ref{fig4}(d). Our numerical results are qualitatively aligned with the conclusions of the works \cite{xingyao2024,2025arXivxinglei}, which demonstrate that a large quantum metric can induce long-range decay for topological edge states. Additionally, other model parameters, such as $m_1$ and $\lambda$, can also tune the quantum metric, thereby influencing the localization properties of the end states (see Appendix \ref{Appendix A}).

\section{Discussion and conclusion }
\label{VI}
Our method utilizes the band inversion between the dispersive and flat bands to modulate the quantum metric. This scenario can be realized in solid state materials, such as electric field-induced band inversion between the first and second Moiré bands in MoTe$_2$/WSe$_2$
heterobilayers \cite{Zhang2021,Symmetricluo}, as well as the band inversion between heavy and light electrons in the topological Kondo insulator SmB$_6$ \cite{PhysRevLett.104.106408,Neupane2013}. In these materials, we anticipate the realization of enhanced quantum metric, which provides a potential platform for investigating quantum metric effects. Furthermore, given the experimental simulation of topological phases in various metamaterials, including acoustic \cite{Xue2022}, photonic \cite{OzawaTomoki2019}, and electrical circuit systems \cite{2024arXiv240909919C}, we believe that these metamaterials also have significant potential for simulating our theoretically proposed models, thus broadening the scope of quantum metric studies in engineered systems.

In summary, we propose a unified framework for engineering topological bands with tunable quantum metric based on coupled “bilayer” Dirac models. The localization properties of gapless boundary states are explicitly explored using a “Bilayer” SSH model. Our work establishes a foundation for further studies on the interplay between band topology and quantum metric.

\section{Acknowledgment}
K.T.L acknowledges the support from the Ministry of Science and Technology, China, and Hong Kong Research Grant Council through Grants No. 2020YFA0309600, No. RFS2021-6S03,
 No. C6025-19G, No. AoE/P-701/20, No. 16310520,
 No. 16307622, and No. 16309223.

\begin{figure}[t]
\centering
\includegraphics[width=3.4in]{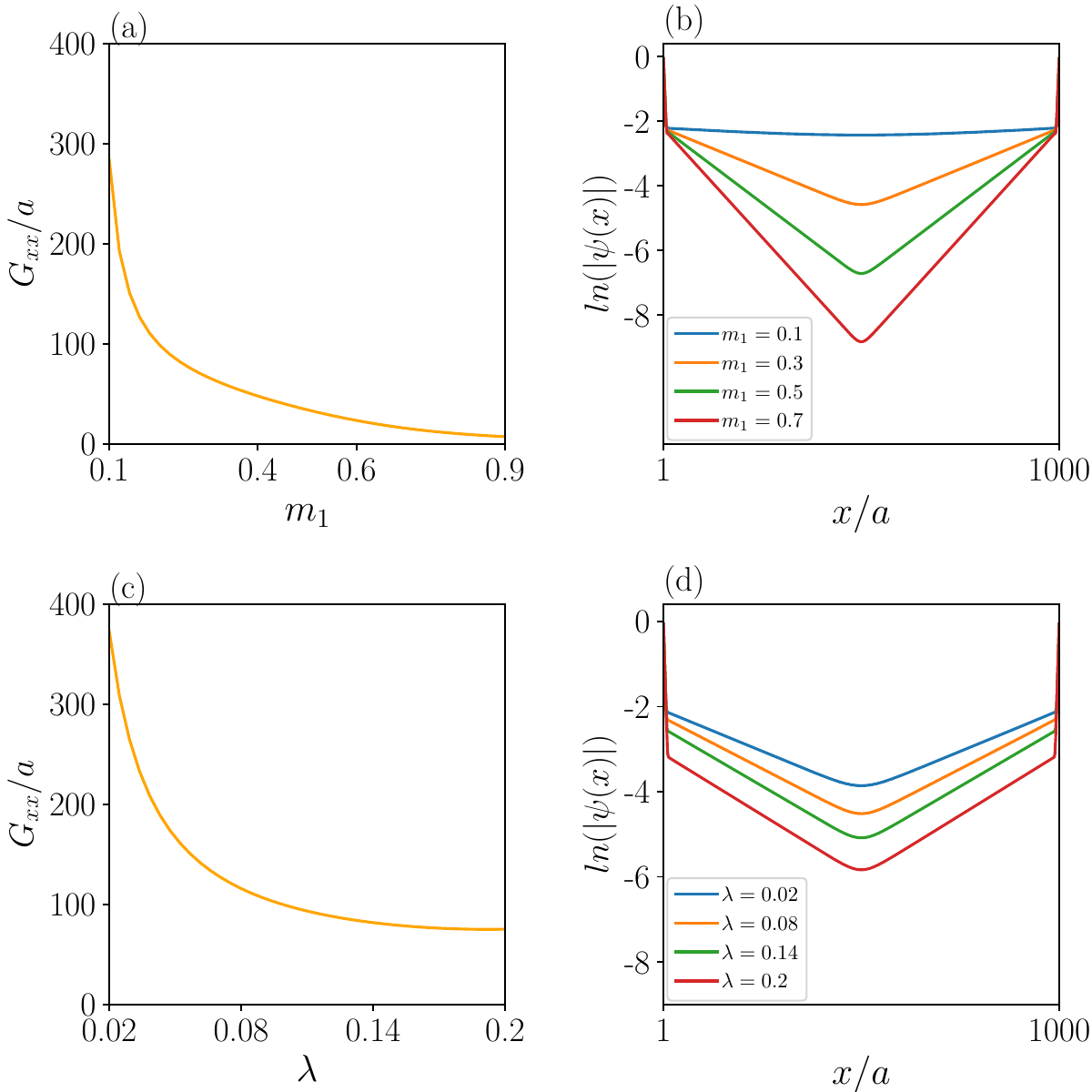}
\caption{Quantum weight and edge states plot for the “Bilayer” SSH model. (a) The plot of quantum weight $G_{xx}$ as a function of $m_1$. (b) The plot of $\text{ln}|\psi(x)|$ for the end states under different setting of $m_1$.
(c) The plot of quantum weight $G_{xx}$ as a function of $\lambda$. (d) The plot of $\text{ln}|\psi(x)|$ for the end states under different setting of $\lambda$.
In (a) and (b), $m_2=0.6$, $\epsilon_1=20$, $\lambda=0.2$, and $\epsilon_2=-0.4$. In (c) and (d), $m_2=0.6$, $\epsilon_1=15$, $\epsilon_2=-0.4$, and $m_1=0.2$.
 }
\label{fig5}
\end{figure}

\appendix
\section{“Bilayer” SSH model}
\label{Appendix A}
The model Hamiltonian of “bilayer” SSH model can be written as 
\beqn
\mathcal{H} = \begin{pmatrix}
\epsilon_1 H_{\text{SSH}}^{(1)} & \lambda  \\
\lambda  & \epsilon_2 H_{\text{SSH}}^{(2)} 
\end{pmatrix},\nonumber\\
H_{\text{SSH}}^{(i)}=M_i(k_x)\sigma_x+\sin k_xa\sigma_y,
\eeqn
where $M_i(k_x)=m_i/\epsilon_i+(1+\cos k_x a)$ for $i=1,2$. The quantum weight $G_{xx}$ of $\mathcal{H}$ can be tuned by model parameters $m_1$ and $\lambda$, as depicted in Figs.~\ref{fig5}(a) and \ref{fig5}(c), respectively. As $m_1$ or $\lambda$ increases, $G_{xx}$ decreases. Correspondingly, the end states of $\mathcal{H}$ become more localized, as shown in Figs.~\ref{fig5}(b) and \ref{fig5}(d).

\bibliography{reference}

\begin{thebibliography}{82}%
\makeatletter
\providecommand \@ifxundefined [1]{%
 \@ifx{#1\undefined}
}%
\providecommand \@ifnum [1]{%
 \ifnum #1\expandafter \@firstoftwo
 \else \expandafter \@secondoftwo
 \fi
}%
\providecommand \@ifx [1]{%
 \ifx #1\expandafter \@firstoftwo
 \else \expandafter \@secondoftwo
 \fi
}%
\providecommand \natexlab [1]{#1}%
\providecommand \enquote  [1]{``#1''}%
\providecommand \bibnamefont  [1]{#1}%
\providecommand \bibfnamefont [1]{#1}%
\providecommand \citenamefont [1]{#1}%
\providecommand \href@noop [0]{\@secondoftwo}%
\providecommand \href [0]{\begingroup \@sanitize@url \@href}%
\providecommand \@href[1]{\@@startlink{#1}\@@href}%
\providecommand \@@href[1]{\endgroup#1\@@endlink}%
\providecommand \@sanitize@url [0]{\catcode `\\12\catcode `\$12\catcode `\&12\catcode `\#12\catcode `\^12\catcode `\_12\catcode `\%12\relax}%
\providecommand \@@startlink[1]{}%
\providecommand \@@endlink[0]{}%
\providecommand \url  [0]{\begingroup\@sanitize@url \@url }%
\providecommand \@url [1]{\endgroup\@href {#1}{\urlprefix }}%
\providecommand \urlprefix  [0]{URL }%
\providecommand \Eprint [0]{\href }%
\providecommand \doibase [0]{https://doi.org/}%
\providecommand \selectlanguage [0]{\@gobble}%
\providecommand \bibinfo  [0]{\@secondoftwo}%
\providecommand \bibfield  [0]{\@secondoftwo}%
\providecommand \translation [1]{[#1]}%
\providecommand \BibitemOpen [0]{}%
\providecommand \bibitemStop [0]{}%
\providecommand \bibitemNoStop [0]{.\EOS\space}%
\providecommand \EOS [0]{\spacefactor3000\relax}%
\providecommand \BibitemShut  [1]{\csname bibitem#1\endcsname}%
\let\auto@bib@innerbib\@empty
\bibitem [{\citenamefont {{Cheng}}(2010)}]{2010arXiv1012.1337C}%
  \BibitemOpen
  \bibfield  {author} {\bibinfo {author} {\bibfnamefont {R.}~\bibnamefont {{Cheng}}},\ }\bibfield  {title} {\bibinfo {title} {{Quantum Geometric Tensor (Fubini-Study Metric) in Simple Quantum System: A pedagogical Introduction}},\ }\href {https://arxiv.org/abs/1012.1337} {\bibfield  {journal} {\bibinfo  {journal} {arXiv:1012.1337}\ } (\bibinfo {year} {2010})}\BibitemShut {NoStop}%
\bibitem [{\citenamefont {Liu}\ \emph {et~al.}(2024)\citenamefont {Liu}, \citenamefont {Qiang}, \citenamefont {Lu},\ and\ \citenamefont {Xie}}]{10.1093/nsr/nwae334}%
  \BibitemOpen
  \bibfield  {author} {\bibinfo {author} {\bibfnamefont {T.}~\bibnamefont {Liu}}, \bibinfo {author} {\bibfnamefont {X.-B.}\ \bibnamefont {Qiang}}, \bibinfo {author} {\bibfnamefont {H.-Z.}\ \bibnamefont {Lu}},\ and\ \bibinfo {author} {\bibfnamefont {X.~C.}\ \bibnamefont {Xie}},\ }\bibfield  {title} {\bibinfo {title} {Quantum geometry in condensed matter},\ }\href {https://doi.org/10.1093/nsr/nwae334} {\bibfield  {journal} {\bibinfo  {journal} {National Science Review}\ }\textbf {\bibinfo {volume} {12}},\ \bibinfo {pages} {nwae334} (\bibinfo {year} {2024})}\BibitemShut {NoStop}%
\bibitem [{\citenamefont {{Yu}}\ \emph {et~al.}(2024)\citenamefont {{Yu}}, \citenamefont {{Bernevig}}, \citenamefont {{Queiroz}}, \citenamefont {{Rossi}}, \citenamefont {{T{\"o}rm{\"a}}},\ and\ \citenamefont {{Yang}}}]{2025arXiv250100098Y}%
  \BibitemOpen
  \bibfield  {author} {\bibinfo {author} {\bibfnamefont {J.}~\bibnamefont {{Yu}}}, \bibinfo {author} {\bibfnamefont {B.~A.}\ \bibnamefont {{Bernevig}}}, \bibinfo {author} {\bibfnamefont {R.}~\bibnamefont {{Queiroz}}}, \bibinfo {author} {\bibfnamefont {E.}~\bibnamefont {{Rossi}}}, \bibinfo {author} {\bibfnamefont {P.}~\bibnamefont {{T{\"o}rm{\"a}}}},\ and\ \bibinfo {author} {\bibfnamefont {B.-J.}\ \bibnamefont {{Yang}}},\ }\bibfield  {title} {\bibinfo {title} {{Quantum Geometry in Quantum Materials}},\ }\href {https://arxiv.org/abs/2501.00098} {\bibfield  {journal} {\bibinfo  {journal} {arXiv:2501.00098}\ } (\bibinfo {year} {2024})}\BibitemShut {NoStop}%
\bibitem [{\citenamefont {Xiao}\ \emph {et~al.}(2010)\citenamefont {Xiao}, \citenamefont {Chang},\ and\ \citenamefont {Niu}}]{RevModPhys.82.1959}%
  \BibitemOpen
  \bibfield  {author} {\bibinfo {author} {\bibfnamefont {D.}~\bibnamefont {Xiao}}, \bibinfo {author} {\bibfnamefont {M.-C.}\ \bibnamefont {Chang}},\ and\ \bibinfo {author} {\bibfnamefont {Q.}~\bibnamefont {Niu}},\ }\bibfield  {title} {\bibinfo {title} {Berry phase effects on electronic properties},\ }\href {https://doi.org/10.1103/RevModPhys.82.1959} {\bibfield  {journal} {\bibinfo  {journal} {Rev. Mod. Phys.}\ }\textbf {\bibinfo {volume} {82}},\ \bibinfo {pages} {1959} (\bibinfo {year} {2010})}\BibitemShut {NoStop}%
\bibitem [{\citenamefont {Hasan}\ and\ \citenamefont {Kane}(2010)}]{RevModPhys.82.3045}%
  \BibitemOpen
  \bibfield  {author} {\bibinfo {author} {\bibfnamefont {M.~Z.}\ \bibnamefont {Hasan}}\ and\ \bibinfo {author} {\bibfnamefont {C.~L.}\ \bibnamefont {Kane}},\ }\bibfield  {title} {\bibinfo {title} {Colloquium: Topological insulators},\ }\href {https://doi.org/10.1103/RevModPhys.82.3045} {\bibfield  {journal} {\bibinfo  {journal} {Rev. Mod. Phys.}\ }\textbf {\bibinfo {volume} {82}},\ \bibinfo {pages} {3045} (\bibinfo {year} {2010})}\BibitemShut {NoStop}%
\bibitem [{\citenamefont {{Qi}}\ and\ \citenamefont {{Zhang}}(2011)}]{Qi2011}%
  \BibitemOpen
  \bibfield  {author} {\bibinfo {author} {\bibfnamefont {X.-L.}\ \bibnamefont {{Qi}}}\ and\ \bibinfo {author} {\bibfnamefont {S.-C.}\ \bibnamefont {{Zhang}}},\ }\bibfield  {title} {\bibinfo {title} {{Topological insulators and superconductors}},\ }\href {https://doi.org/10.1103/RevModPhys.83.1057} {\bibfield  {journal} {\bibinfo  {journal} {Reviews of Modern Physics}\ }\textbf {\bibinfo {volume} {83}},\ \bibinfo {pages} {1057} (\bibinfo {year} {2011})}\BibitemShut {NoStop}%
\bibitem [{\citenamefont {Chiu}\ \emph {et~al.}(2016)\citenamefont {Chiu}, \citenamefont {Teo}, \citenamefont {Schnyder},\ and\ \citenamefont {Ryu}}]{Chiu2016}%
  \BibitemOpen
  \bibfield  {author} {\bibinfo {author} {\bibfnamefont {C.-K.}\ \bibnamefont {Chiu}}, \bibinfo {author} {\bibfnamefont {J.~C.~Y.}\ \bibnamefont {Teo}}, \bibinfo {author} {\bibfnamefont {A.~P.}\ \bibnamefont {Schnyder}},\ and\ \bibinfo {author} {\bibfnamefont {S.}~\bibnamefont {Ryu}},\ }\bibfield  {title} {\bibinfo {title} {Classification of topological quantum matter with symmetries},\ }\href {https://doi.org/10.1103/RevModPhys.88.035005} {\bibfield  {journal} {\bibinfo  {journal} {Rev. Mod. Phys.}\ }\textbf {\bibinfo {volume} {88}},\ \bibinfo {pages} {035005} (\bibinfo {year} {2016})}\BibitemShut {NoStop}%
\bibitem [{\citenamefont {Kane}\ and\ \citenamefont {Mele}(2005)}]{Kane2005}%
  \BibitemOpen
  \bibfield  {author} {\bibinfo {author} {\bibfnamefont {C.~L.}\ \bibnamefont {Kane}}\ and\ \bibinfo {author} {\bibfnamefont {E.~J.}\ \bibnamefont {Mele}},\ }\bibfield  {title} {\bibinfo {title} {Quantum spin hall effect in graphene},\ }\href {https://doi.org/10.1103/PhysRevLett.95.226801} {\bibfield  {journal} {\bibinfo  {journal} {Phys. Rev. Lett.}\ }\textbf {\bibinfo {volume} {95}},\ \bibinfo {pages} {226801} (\bibinfo {year} {2005})}\BibitemShut {NoStop}%
\bibitem [{\citenamefont {Bernevig}\ \emph {et~al.}(2006)\citenamefont {Bernevig}, \citenamefont {Hughes},\ and\ \citenamefont {Zhang}}]{Bernevig2006}%
  \BibitemOpen
  \bibfield  {author} {\bibinfo {author} {\bibfnamefont {B.~A.}\ \bibnamefont {Bernevig}}, \bibinfo {author} {\bibfnamefont {T.~L.}\ \bibnamefont {Hughes}},\ and\ \bibinfo {author} {\bibfnamefont {S.-C.}\ \bibnamefont {Zhang}},\ }\bibfield  {title} {\bibinfo {title} {Quantum spin hall effect and topological phase transition in hgte quantum wells},\ }\href {https://doi.org/10.1126/science.1133734} {\bibfield  {journal} {\bibinfo  {journal} {Science}\ }\textbf {\bibinfo {volume} {314}},\ \bibinfo {pages} {1757} (\bibinfo {year} {2006})}\BibitemShut {NoStop}%
\bibitem [{\citenamefont {Fu}\ \emph {et~al.}(2007{\natexlab{a}})\citenamefont {Fu}, \citenamefont {Kane},\ and\ \citenamefont {Mele}}]{PhysRevLett.98.106803}%
  \BibitemOpen
  \bibfield  {author} {\bibinfo {author} {\bibfnamefont {L.}~\bibnamefont {Fu}}, \bibinfo {author} {\bibfnamefont {C.~L.}\ \bibnamefont {Kane}},\ and\ \bibinfo {author} {\bibfnamefont {E.~J.}\ \bibnamefont {Mele}},\ }\bibfield  {title} {\bibinfo {title} {Topological insulators in three dimensions},\ }\href {https://doi.org/10.1103/PhysRevLett.98.106803} {\bibfield  {journal} {\bibinfo  {journal} {Phys. Rev. Lett.}\ }\textbf {\bibinfo {volume} {98}},\ \bibinfo {pages} {106803} (\bibinfo {year} {2007}{\natexlab{a}})}\BibitemShut {NoStop}%
\bibitem [{\citenamefont {Fu}\ and\ \citenamefont {Kane}(2008)}]{Fu2008}%
  \BibitemOpen
  \bibfield  {author} {\bibinfo {author} {\bibfnamefont {L.}~\bibnamefont {Fu}}\ and\ \bibinfo {author} {\bibfnamefont {C.~L.}\ \bibnamefont {Kane}},\ }\bibfield  {title} {\bibinfo {title} {Superconducting proximity effect and majorana fermions at the surface of a topological insulator},\ }\href {https://doi.org/10.1103/PhysRevLett.100.096407} {\bibfield  {journal} {\bibinfo  {journal} {Phys. Rev. Lett.}\ }\textbf {\bibinfo {volume} {100}},\ \bibinfo {pages} {096407} (\bibinfo {year} {2008})}\BibitemShut {NoStop}%
\bibitem [{\citenamefont {Lutchyn}\ \emph {et~al.}(2010)\citenamefont {Lutchyn}, \citenamefont {Sau},\ and\ \citenamefont {Das~Sarma}}]{Lutchyn2010}%
  \BibitemOpen
  \bibfield  {author} {\bibinfo {author} {\bibfnamefont {R.~M.}\ \bibnamefont {Lutchyn}}, \bibinfo {author} {\bibfnamefont {J.~D.}\ \bibnamefont {Sau}},\ and\ \bibinfo {author} {\bibfnamefont {S.}~\bibnamefont {Das~Sarma}},\ }\bibfield  {title} {\bibinfo {title} {Majorana fermions and a topological phase transition in semiconductor-superconductor heterostructures},\ }\href {https://doi.org/10.1103/PhysRevLett.105.077001} {\bibfield  {journal} {\bibinfo  {journal} {Phys. Rev. Lett.}\ }\textbf {\bibinfo {volume} {105}},\ \bibinfo {pages} {077001} (\bibinfo {year} {2010})}\BibitemShut {NoStop}%
\bibitem [{\citenamefont {Luo}\ \emph {et~al.}(2025)\citenamefont {Luo}, \citenamefont {Li}, \citenamefont {Xiao},\ and\ \citenamefont {Wu}}]{Luo20241}%
  \BibitemOpen
  \bibfield  {author} {\bibinfo {author} {\bibfnamefont {X.-J.}\ \bibnamefont {Luo}}, \bibinfo {author} {\bibfnamefont {J.-Z.}\ \bibnamefont {Li}}, \bibinfo {author} {\bibfnamefont {M.}~\bibnamefont {Xiao}},\ and\ \bibinfo {author} {\bibfnamefont {F.}~\bibnamefont {Wu}},\ }\bibfield  {title} {\bibinfo {title} {Characterization of higher-order topological superconductors using bott indices},\ }\href {https://doi.org/10.1103/PhysRevB.111.184516} {\bibfield  {journal} {\bibinfo  {journal} {Phys. Rev. B}\ }\textbf {\bibinfo {volume} {111}},\ \bibinfo {pages} {184516} (\bibinfo {year} {2025})}\BibitemShut {NoStop}%
\bibitem [{\citenamefont {Benalcazar}\ \emph {et~al.}(2017{\natexlab{a}})\citenamefont {Benalcazar}, \citenamefont {Bernevig},\ and\ \citenamefont {Hughes}}]{Benalcazar2017a}%
  \BibitemOpen
  \bibfield  {author} {\bibinfo {author} {\bibfnamefont {W.~A.}\ \bibnamefont {Benalcazar}}, \bibinfo {author} {\bibfnamefont {B.~A.}\ \bibnamefont {Bernevig}},\ and\ \bibinfo {author} {\bibfnamefont {T.~L.}\ \bibnamefont {Hughes}},\ }\bibfield  {title} {\bibinfo {title} {Quantized electric multipole insulators},\ }\href {https://doi.org/10.1126/science.aah6442} {\bibfield  {journal} {\bibinfo  {journal} {Science}\ }\textbf {\bibinfo {volume} {357}},\ \bibinfo {pages} {61} (\bibinfo {year} {2017}{\natexlab{a}})}\BibitemShut {NoStop}%
\bibitem [{\citenamefont {Benalcazar}\ \emph {et~al.}(2017{\natexlab{b}})\citenamefont {Benalcazar}, \citenamefont {Bernevig},\ and\ \citenamefont {Hughes}}]{Benalcazar2017}%
  \BibitemOpen
  \bibfield  {author} {\bibinfo {author} {\bibfnamefont {W.~A.}\ \bibnamefont {Benalcazar}}, \bibinfo {author} {\bibfnamefont {B.~A.}\ \bibnamefont {Bernevig}},\ and\ \bibinfo {author} {\bibfnamefont {T.~L.}\ \bibnamefont {Hughes}},\ }\bibfield  {title} {\bibinfo {title} {Electric multipole moments, topological multipole moment pumping, and chiral hinge states in crystalline insulators},\ }\href {https://doi.org/10.1103/PhysRevB.96.245115} {\bibfield  {journal} {\bibinfo  {journal} {Phys. Rev. B}\ }\textbf {\bibinfo {volume} {96}},\ \bibinfo {pages} {245115} (\bibinfo {year} {2017}{\natexlab{b}})}\BibitemShut {NoStop}%
\bibitem [{\citenamefont {Song}\ \emph {et~al.}(2017)\citenamefont {Song}, \citenamefont {Fang},\ and\ \citenamefont {Fang}}]{Song2017}%
  \BibitemOpen
  \bibfield  {author} {\bibinfo {author} {\bibfnamefont {Z.}~\bibnamefont {Song}}, \bibinfo {author} {\bibfnamefont {Z.}~\bibnamefont {Fang}},\ and\ \bibinfo {author} {\bibfnamefont {C.}~\bibnamefont {Fang}},\ }\bibfield  {title} {\bibinfo {title} {$(d\ensuremath{-}2)$-dimensional edge states of rotation symmetry protected topological states},\ }\href {https://doi.org/10.1103/PhysRevLett.119.246402} {\bibfield  {journal} {\bibinfo  {journal} {Phys. Rev. Lett.}\ }\textbf {\bibinfo {volume} {119}},\ \bibinfo {pages} {246402} (\bibinfo {year} {2017})}\BibitemShut {NoStop}%
\bibitem [{\citenamefont {Luo}\ \emph {et~al.}(2021)\citenamefont {Luo}, \citenamefont {Pan},\ and\ \citenamefont {Liu}}]{Luo2021a}%
  \BibitemOpen
  \bibfield  {author} {\bibinfo {author} {\bibfnamefont {X.-J.}\ \bibnamefont {Luo}}, \bibinfo {author} {\bibfnamefont {X.-H.}\ \bibnamefont {Pan}},\ and\ \bibinfo {author} {\bibfnamefont {X.}~\bibnamefont {Liu}},\ }\bibfield  {title} {\bibinfo {title} {Higher-order topological superconductors based on weak topological insulators},\ }\href {https://doi.org/10.1103/PhysRevB.104.104510} {\bibfield  {journal} {\bibinfo  {journal} {Phys. Rev. B}\ }\textbf {\bibinfo {volume} {104}},\ \bibinfo {pages} {104510} (\bibinfo {year} {2021})}\BibitemShut {NoStop}%
\bibitem [{\citenamefont {Luo}\ and\ \citenamefont {Wu}(2023)}]{luo2024}%
  \BibitemOpen
  \bibfield  {author} {\bibinfo {author} {\bibfnamefont {X.-J.}\ \bibnamefont {Luo}}\ and\ \bibinfo {author} {\bibfnamefont {F.}~\bibnamefont {Wu}},\ }\bibfield  {title} {\bibinfo {title} {Generalization of benalcazar-bernevig-hughes model to arbitrary dimensions},\ }\href {https://doi.org/10.1103/PhysRevB.108.075143} {\bibfield  {journal} {\bibinfo  {journal} {Phys. Rev. B}\ }\textbf {\bibinfo {volume} {108}},\ \bibinfo {pages} {075143} (\bibinfo {year} {2023})}\BibitemShut {NoStop}%
\bibitem [{\citenamefont {Luo}\ \emph {et~al.}(2023{\natexlab{a}})\citenamefont {Luo}, \citenamefont {Pan}, \citenamefont {Liu},\ and\ \citenamefont {Liu}}]{Luo2023a}%
  \BibitemOpen
  \bibfield  {author} {\bibinfo {author} {\bibfnamefont {X.-J.}\ \bibnamefont {Luo}}, \bibinfo {author} {\bibfnamefont {X.-H.}\ \bibnamefont {Pan}}, \bibinfo {author} {\bibfnamefont {C.-X.}\ \bibnamefont {Liu}},\ and\ \bibinfo {author} {\bibfnamefont {X.}~\bibnamefont {Liu}},\ }\bibfield  {title} {\bibinfo {title} {Higher-order topological phases emerging from su-schrieffer-heeger stacking},\ }\href {https://doi.org/10.1103/PhysRevB.107.045118} {\bibfield  {journal} {\bibinfo  {journal} {Phys. Rev. B}\ }\textbf {\bibinfo {volume} {107}},\ \bibinfo {pages} {045118} (\bibinfo {year} {2023}{\natexlab{a}})}\BibitemShut {NoStop}%
\bibitem [{\citenamefont {Gao}\ \emph {et~al.}(2014)\citenamefont {Gao}, \citenamefont {Yang},\ and\ \citenamefont {Niu}}]{gaoyang2014}%
  \BibitemOpen
  \bibfield  {author} {\bibinfo {author} {\bibfnamefont {Y.}~\bibnamefont {Gao}}, \bibinfo {author} {\bibfnamefont {S.~A.}\ \bibnamefont {Yang}},\ and\ \bibinfo {author} {\bibfnamefont {Q.}~\bibnamefont {Niu}},\ }\bibfield  {title} {\bibinfo {title} {Field induced positional shift of bloch electrons and its dynamical implications},\ }\href {https://doi.org/10.1103/PhysRevLett.112.166601} {\bibfield  {journal} {\bibinfo  {journal} {Phys. Rev. Lett.}\ }\textbf {\bibinfo {volume} {112}},\ \bibinfo {pages} {166601} (\bibinfo {year} {2014})}\BibitemShut {NoStop}%
\bibitem [{\citenamefont {Liu}\ \emph {et~al.}(2021)\citenamefont {Liu}, \citenamefont {Zhao}, \citenamefont {Huang}, \citenamefont {Wu}, \citenamefont {Sheng}, \citenamefont {Xiao},\ and\ \citenamefont {Yang}}]{LiuHuiying2021}%
  \BibitemOpen
  \bibfield  {author} {\bibinfo {author} {\bibfnamefont {H.}~\bibnamefont {Liu}}, \bibinfo {author} {\bibfnamefont {J.}~\bibnamefont {Zhao}}, \bibinfo {author} {\bibfnamefont {Y.-X.}\ \bibnamefont {Huang}}, \bibinfo {author} {\bibfnamefont {W.}~\bibnamefont {Wu}}, \bibinfo {author} {\bibfnamefont {X.-L.}\ \bibnamefont {Sheng}}, \bibinfo {author} {\bibfnamefont {C.}~\bibnamefont {Xiao}},\ and\ \bibinfo {author} {\bibfnamefont {S.~A.}\ \bibnamefont {Yang}},\ }\bibfield  {title} {\bibinfo {title} {Intrinsic second-order anomalous hall effect and its application in compensated antiferromagnets},\ }\href {https://doi.org/10.1103/PhysRevLett.127.277202} {\bibfield  {journal} {\bibinfo  {journal} {Phys. Rev. Lett.}\ }\textbf {\bibinfo {volume} {127}},\ \bibinfo {pages} {277202} (\bibinfo {year} {2021})}\BibitemShut {NoStop}%
\bibitem [{\citenamefont {Wang}\ \emph {et~al.}(2021)\citenamefont {Wang}, \citenamefont {Gao},\ and\ \citenamefont {Xiao}}]{wangchong2021}%
  \BibitemOpen
  \bibfield  {author} {\bibinfo {author} {\bibfnamefont {C.}~\bibnamefont {Wang}}, \bibinfo {author} {\bibfnamefont {Y.}~\bibnamefont {Gao}},\ and\ \bibinfo {author} {\bibfnamefont {D.}~\bibnamefont {Xiao}},\ }\bibfield  {title} {\bibinfo {title} {Intrinsic nonlinear hall effect in antiferromagnetic tetragonal cumnas},\ }\href {https://doi.org/10.1103/PhysRevLett.127.277201} {\bibfield  {journal} {\bibinfo  {journal} {Phys. Rev. Lett.}\ }\textbf {\bibinfo {volume} {127}},\ \bibinfo {pages} {277201} (\bibinfo {year} {2021})}\BibitemShut {NoStop}%
\bibitem [{\citenamefont {Gao}\ \emph {et~al.}(2023)\citenamefont {Gao}, \citenamefont {Liu}, \citenamefont {Qiu}, \citenamefont {Ghosh}, \citenamefont {Trevisan}, \citenamefont {Onishi}, \citenamefont {Hu}, \citenamefont {Qian}, \citenamefont {Tien}, \citenamefont {Chen}, \citenamefont {Huang}, \citenamefont {Bérubé}, \citenamefont {Li}, \citenamefont {Tzschaschel}, \citenamefont {Dinh}, \citenamefont {Sun}, \citenamefont {Ho}, \citenamefont {Lien}, \citenamefont {Singh}, \citenamefont {Watanabe}, \citenamefont {Taniguchi}, \citenamefont {Bell}, \citenamefont {Lin}, \citenamefont {Chang}, \citenamefont {Du}, \citenamefont {Bansil}, \citenamefont {Fu}, \citenamefont {Ni}, \citenamefont {Orth}, \citenamefont {Ma},\ and\ \citenamefont {Xu}}]{anyuanguo2023}%
  \BibitemOpen
  \bibfield  {author} {\bibinfo {author} {\bibfnamefont {A.}~\bibnamefont {Gao}}, \bibinfo {author} {\bibfnamefont {Y.-F.}\ \bibnamefont {Liu}}, \bibinfo {author} {\bibfnamefont {J.-X.}\ \bibnamefont {Qiu}}, \bibinfo {author} {\bibfnamefont {B.}~\bibnamefont {Ghosh}}, \bibinfo {author} {\bibfnamefont {T.~V.}\ \bibnamefont {Trevisan}}, \bibinfo {author} {\bibfnamefont {Y.}~\bibnamefont {Onishi}}, \bibinfo {author} {\bibfnamefont {C.}~\bibnamefont {Hu}}, \bibinfo {author} {\bibfnamefont {T.}~\bibnamefont {Qian}}, \bibinfo {author} {\bibfnamefont {H.-J.}\ \bibnamefont {Tien}}, \bibinfo {author} {\bibfnamefont {S.-W.}\ \bibnamefont {Chen}}, \bibinfo {author} {\bibfnamefont {M.}~\bibnamefont {Huang}}, \bibinfo {author} {\bibfnamefont {D.}~\bibnamefont {Bérubé}}, \bibinfo {author} {\bibfnamefont {H.}~\bibnamefont {Li}}, \bibinfo {author} {\bibfnamefont {C.}~\bibnamefont {Tzschaschel}}, \bibinfo {author} {\bibfnamefont {T.}~\bibnamefont {Dinh}}, \bibinfo {author} {\bibfnamefont {Z.}~\bibnamefont {Sun}}, \bibinfo
  {author} {\bibfnamefont {S.-C.}\ \bibnamefont {Ho}}, \bibinfo {author} {\bibfnamefont {S.-W.}\ \bibnamefont {Lien}}, \bibinfo {author} {\bibfnamefont {B.}~\bibnamefont {Singh}}, \bibinfo {author} {\bibfnamefont {K.}~\bibnamefont {Watanabe}}, \bibinfo {author} {\bibfnamefont {T.}~\bibnamefont {Taniguchi}}, \bibinfo {author} {\bibfnamefont {D.~C.}\ \bibnamefont {Bell}}, \bibinfo {author} {\bibfnamefont {H.}~\bibnamefont {Lin}}, \bibinfo {author} {\bibfnamefont {T.-R.}\ \bibnamefont {Chang}}, \bibinfo {author} {\bibfnamefont {C.~R.}\ \bibnamefont {Du}}, \bibinfo {author} {\bibfnamefont {A.}~\bibnamefont {Bansil}}, \bibinfo {author} {\bibfnamefont {L.}~\bibnamefont {Fu}}, \bibinfo {author} {\bibfnamefont {N.}~\bibnamefont {Ni}}, \bibinfo {author} {\bibfnamefont {P.~P.}\ \bibnamefont {Orth}}, \bibinfo {author} {\bibfnamefont {Q.}~\bibnamefont {Ma}},\ and\ \bibinfo {author} {\bibfnamefont {S.-Y.}\ \bibnamefont {Xu}},\ }\bibfield  {title} {\bibinfo {title} {Quantum metric nonlinear hall effect in a topological
  antiferromagnetic heterostructure},\ }\href {https://doi.org/10.1126/science.adf1506} {\bibfield  {journal} {\bibinfo  {journal} {Science}\ }\textbf {\bibinfo {volume} {381}},\ \bibinfo {pages} {181} (\bibinfo {year} {2023})}\BibitemShut {NoStop}%
\bibitem [{\citenamefont {Souza}\ \emph {et~al.}(2000)\citenamefont {Souza}, \citenamefont {Wilkens},\ and\ \citenamefont {Martin}}]{SouzaIvo2000}%
  \BibitemOpen
  \bibfield  {author} {\bibinfo {author} {\bibfnamefont {I.}~\bibnamefont {Souza}}, \bibinfo {author} {\bibfnamefont {T.}~\bibnamefont {Wilkens}},\ and\ \bibinfo {author} {\bibfnamefont {R.~M.}\ \bibnamefont {Martin}},\ }\bibfield  {title} {\bibinfo {title} {Polarization and localization in insulators: Generating function approach},\ }\href {https://doi.org/10.1103/PhysRevB.62.1666} {\bibfield  {journal} {\bibinfo  {journal} {Phys. Rev. B}\ }\textbf {\bibinfo {volume} {62}},\ \bibinfo {pages} {1666} (\bibinfo {year} {2000})}\BibitemShut {NoStop}%
\bibitem [{\citenamefont {Gao}\ and\ \citenamefont {Xiao}(2019)}]{GaoYang2019}%
  \BibitemOpen
  \bibfield  {author} {\bibinfo {author} {\bibfnamefont {Y.}~\bibnamefont {Gao}}\ and\ \bibinfo {author} {\bibfnamefont {D.}~\bibnamefont {Xiao}},\ }\bibfield  {title} {\bibinfo {title} {Nonreciprocal directional dichroism induced by the quantum metric dipole},\ }\href {https://doi.org/10.1103/PhysRevLett.122.227402} {\bibfield  {journal} {\bibinfo  {journal} {Phys. Rev. Lett.}\ }\textbf {\bibinfo {volume} {122}},\ \bibinfo {pages} {227402} (\bibinfo {year} {2019})}\BibitemShut {NoStop}%
\bibitem [{\citenamefont {Jankowski}\ and\ \citenamefont {Slager}(2024)}]{Jankowski2024}%
  \BibitemOpen
  \bibfield  {author} {\bibinfo {author} {\bibfnamefont {W.~J.}\ \bibnamefont {Jankowski}}\ and\ \bibinfo {author} {\bibfnamefont {R.-J.}\ \bibnamefont {Slager}},\ }\bibfield  {title} {\bibinfo {title} {Quantized integrated shift effect in multigap topological phases},\ }\href {https://doi.org/10.1103/PhysRevLett.133.186601} {\bibfield  {journal} {\bibinfo  {journal} {Phys. Rev. Lett.}\ }\textbf {\bibinfo {volume} {133}},\ \bibinfo {pages} {186601} (\bibinfo {year} {2024})}\BibitemShut {NoStop}%
\bibitem [{\citenamefont {Onishi}\ and\ \citenamefont {Fu}(2024)}]{Onishi2024}%
  \BibitemOpen
  \bibfield  {author} {\bibinfo {author} {\bibfnamefont {Y.}~\bibnamefont {Onishi}}\ and\ \bibinfo {author} {\bibfnamefont {L.}~\bibnamefont {Fu}},\ }\bibfield  {title} {\bibinfo {title} {Fundamental bound on topological gap},\ }\href {https://doi.org/10.1103/PhysRevX.14.011052} {\bibfield  {journal} {\bibinfo  {journal} {Phys. Rev. X}\ }\textbf {\bibinfo {volume} {14}},\ \bibinfo {pages} {011052} (\bibinfo {year} {2024})}\BibitemShut {NoStop}%
\bibitem [{\citenamefont {Komissarov}\ \emph {et~al.}(2024)\citenamefont {Komissarov}, \citenamefont {Holder},\ and\ \citenamefont {Queiroz}}]{Komissarov2024}%
  \BibitemOpen
  \bibfield  {author} {\bibinfo {author} {\bibfnamefont {I.}~\bibnamefont {Komissarov}}, \bibinfo {author} {\bibfnamefont {T.}~\bibnamefont {Holder}},\ and\ \bibinfo {author} {\bibfnamefont {R.}~\bibnamefont {Queiroz}},\ }\bibfield  {title} {\bibinfo {title} {The quantum geometric origin of capacitance in insulators},\ }\href {https://doi.org/10.1038/s41467-024-48808-x} {\bibfield  {journal} {\bibinfo  {journal} {Nature Communications}\ }\textbf {\bibinfo {volume} {15}},\ \bibinfo {pages} {4621} (\bibinfo {year} {2024})}\BibitemShut {NoStop}%
\bibitem [{\citenamefont {Verma}\ and\ \citenamefont {Queiroz}(2025)}]{Verma2024step}%
  \BibitemOpen
  \bibfield  {author} {\bibinfo {author} {\bibfnamefont {N.}~\bibnamefont {Verma}}\ and\ \bibinfo {author} {\bibfnamefont {R.}~\bibnamefont {Queiroz}},\ }\bibfield  {title} {\bibinfo {title} {Framework to measure quantum metric from step response},\ }\href {https://doi.org/10.1103/PhysRevLett.134.106403} {\bibfield  {journal} {\bibinfo  {journal} {Phys. Rev. Lett.}\ }\textbf {\bibinfo {volume} {134}},\ \bibinfo {pages} {106403} (\bibinfo {year} {2025})}\BibitemShut {NoStop}%
\bibitem [{\citenamefont {{Verma}}\ and\ \citenamefont {{Queiroz}}(2024)}]{Verma2024a}%
  \BibitemOpen
  \bibfield  {author} {\bibinfo {author} {\bibfnamefont {N.}~\bibnamefont {{Verma}}}\ and\ \bibinfo {author} {\bibfnamefont {R.}~\bibnamefont {{Queiroz}}},\ }\bibfield  {title} {\bibinfo {title} {{Instantaneous Response and Quantum Geometry of Insulators}},\ }\href {https://arxiv.org/abs/2403.07052} {\bibfield  {journal} {\bibinfo  {journal} {arXiv:2403.07052}\ } (\bibinfo {year} {2024})}\BibitemShut {NoStop}%
\bibitem [{\citenamefont {Resta}(2025)}]{Resta2025}%
  \BibitemOpen
  \bibfield  {author} {\bibinfo {author} {\bibfnamefont {R.}~\bibnamefont {Resta}},\ }\bibfield  {title} {\bibinfo {title} {Nonadiabatic quantum geometry and optical conductivity},\ }\href {https://doi.org/10.1103/PhysRevB.111.205107} {\bibfield  {journal} {\bibinfo  {journal} {Phys. Rev. B}\ }\textbf {\bibinfo {volume} {111}},\ \bibinfo {pages} {205107} (\bibinfo {year} {2025})}\BibitemShut {NoStop}%
\bibitem [{\citenamefont {Paul}(2024)}]{PaulNisarga2024}%
  \BibitemOpen
  \bibfield  {author} {\bibinfo {author} {\bibfnamefont {N.}~\bibnamefont {Paul}},\ }\bibfield  {title} {\bibinfo {title} {Area-law entanglement from quantum geometry},\ }\href {https://doi.org/10.1103/PhysRevB.109.085146} {\bibfield  {journal} {\bibinfo  {journal} {Phys. Rev. B}\ }\textbf {\bibinfo {volume} {109}},\ \bibinfo {pages} {085146} (\bibinfo {year} {2024})}\BibitemShut {NoStop}%
\bibitem [{\citenamefont {Tam}\ \emph {et~al.}(2024)\citenamefont {Tam}, \citenamefont {Herzog-Arbeitman},\ and\ \citenamefont {Yu}}]{TamPokMan2024}%
  \BibitemOpen
  \bibfield  {author} {\bibinfo {author} {\bibfnamefont {P.~M.}\ \bibnamefont {Tam}}, \bibinfo {author} {\bibfnamefont {J.}~\bibnamefont {Herzog-Arbeitman}},\ and\ \bibinfo {author} {\bibfnamefont {J.}~\bibnamefont {Yu}},\ }\bibfield  {title} {\bibinfo {title} {Corner charge fluctuation as an observable for quantum geometry and entanglement in two-dimensional insulators},\ }\href {https://doi.org/10.1103/PhysRevLett.133.246603} {\bibfield  {journal} {\bibinfo  {journal} {Phys. Rev. Lett.}\ }\textbf {\bibinfo {volume} {133}},\ \bibinfo {pages} {246603} (\bibinfo {year} {2024})}\BibitemShut {NoStop}%
\bibitem [{\citenamefont {Zhou}(2024)}]{ZhouLongwen2024}%
  \BibitemOpen
  \bibfield  {author} {\bibinfo {author} {\bibfnamefont {L.}~\bibnamefont {Zhou}},\ }\bibfield  {title} {\bibinfo {title} {Quantum geometry and geometric entanglement entropy of one-dimensional floquet topological matter},\ }\href {https://doi.org/10.1103/PhysRevB.110.054310} {\bibfield  {journal} {\bibinfo  {journal} {Phys. Rev. B}\ }\textbf {\bibinfo {volume} {110}},\ \bibinfo {pages} {054310} (\bibinfo {year} {2024})}\BibitemShut {NoStop}%
\bibitem [{\citenamefont {{Kruchkov}}\ and\ \citenamefont {{Ryu}}(2024)}]{Kruchkov2024}%
  \BibitemOpen
  \bibfield  {author} {\bibinfo {author} {\bibfnamefont {A.}~\bibnamefont {{Kruchkov}}}\ and\ \bibinfo {author} {\bibfnamefont {S.}~\bibnamefont {{Ryu}}},\ }\bibfield  {title} {\bibinfo {title} {{Entanglement entropy in lattice models with quantum metric}},\ }\href {https://arxiv.org/abs/2408.10314} {\bibfield  {journal} {\bibinfo  {journal} {arXiv:2408.10314}\ } (\bibinfo {year} {2024})}\BibitemShut {NoStop}%
\bibitem [{\citenamefont {Wu}\ \emph {et~al.}(2025)\citenamefont {Wu}, \citenamefont {Cai}, \citenamefont {Cheng},\ and\ \citenamefont {Kumar}}]{wuxiaochuan2024}%
  \BibitemOpen
  \bibfield  {author} {\bibinfo {author} {\bibfnamefont {X.-C.}\ \bibnamefont {Wu}}, \bibinfo {author} {\bibfnamefont {K.-L.}\ \bibnamefont {Cai}}, \bibinfo {author} {\bibfnamefont {M.}~\bibnamefont {Cheng}},\ and\ \bibinfo {author} {\bibfnamefont {P.}~\bibnamefont {Kumar}},\ }\bibfield  {title} {\bibinfo {title} {Corner charge fluctuations and many-body quantum geometry},\ }\href {https://doi.org/10.1103/PhysRevB.111.115124} {\bibfield  {journal} {\bibinfo  {journal} {Phys. Rev. B}\ }\textbf {\bibinfo {volume} {111}},\ \bibinfo {pages} {115124} (\bibinfo {year} {2025})}\BibitemShut {NoStop}%
\bibitem [{\citenamefont {Peotta}\ and\ \citenamefont {T{\"o}rm{\"a}}(2015)}]{Peotta2015}%
  \BibitemOpen
  \bibfield  {author} {\bibinfo {author} {\bibfnamefont {S.}~\bibnamefont {Peotta}}\ and\ \bibinfo {author} {\bibfnamefont {P.}~\bibnamefont {T{\"o}rm{\"a}}},\ }\bibfield  {title} {\bibinfo {title} {Superfluidity in topologically nontrivial flat bands},\ }\href {https://doi.org/10.1038/ncomms9944} {\bibfield  {journal} {\bibinfo  {journal} {Nature Communications}\ }\textbf {\bibinfo {volume} {6}},\ \bibinfo {pages} {8944} (\bibinfo {year} {2015})}\BibitemShut {NoStop}%
\bibitem [{\citenamefont {Julku}\ \emph {et~al.}(2016)\citenamefont {Julku}, \citenamefont {Peotta}, \citenamefont {Vanhala}, \citenamefont {Kim},\ and\ \citenamefont {T\"orm\"a}}]{Julku2017}%
  \BibitemOpen
  \bibfield  {author} {\bibinfo {author} {\bibfnamefont {A.}~\bibnamefont {Julku}}, \bibinfo {author} {\bibfnamefont {S.}~\bibnamefont {Peotta}}, \bibinfo {author} {\bibfnamefont {T.~I.}\ \bibnamefont {Vanhala}}, \bibinfo {author} {\bibfnamefont {D.-H.}\ \bibnamefont {Kim}},\ and\ \bibinfo {author} {\bibfnamefont {P.}~\bibnamefont {T\"orm\"a}},\ }\bibfield  {title} {\bibinfo {title} {Geometric origin of superfluidity in the lieb-lattice flat band},\ }\href {https://doi.org/10.1103/PhysRevLett.117.045303} {\bibfield  {journal} {\bibinfo  {journal} {Phys. Rev. Lett.}\ }\textbf {\bibinfo {volume} {117}},\ \bibinfo {pages} {045303} (\bibinfo {year} {2016})}\BibitemShut {NoStop}%
\bibitem [{\citenamefont {Xie}\ \emph {et~al.}(2020)\citenamefont {Xie}, \citenamefont {Song}, \citenamefont {Lian},\ and\ \citenamefont {Bernevig}}]{XieFang2020}%
  \BibitemOpen
  \bibfield  {author} {\bibinfo {author} {\bibfnamefont {F.}~\bibnamefont {Xie}}, \bibinfo {author} {\bibfnamefont {Z.}~\bibnamefont {Song}}, \bibinfo {author} {\bibfnamefont {B.}~\bibnamefont {Lian}},\ and\ \bibinfo {author} {\bibfnamefont {B.~A.}\ \bibnamefont {Bernevig}},\ }\bibfield  {title} {\bibinfo {title} {Topology-bounded superfluid weight in twisted bilayer graphene},\ }\href {https://doi.org/10.1103/PhysRevLett.124.167002} {\bibfield  {journal} {\bibinfo  {journal} {Phys. Rev. Lett.}\ }\textbf {\bibinfo {volume} {124}},\ \bibinfo {pages} {167002} (\bibinfo {year} {2020})}\BibitemShut {NoStop}%
\bibitem [{\citenamefont {Herzog-Arbeitman}\ \emph {et~al.}(2022)\citenamefont {Herzog-Arbeitman}, \citenamefont {Peri}, \citenamefont {Schindler}, \citenamefont {Huber},\ and\ \citenamefont {Bernevig}}]{HerzogArbeitman2022}%
  \BibitemOpen
  \bibfield  {author} {\bibinfo {author} {\bibfnamefont {J.}~\bibnamefont {Herzog-Arbeitman}}, \bibinfo {author} {\bibfnamefont {V.}~\bibnamefont {Peri}}, \bibinfo {author} {\bibfnamefont {F.}~\bibnamefont {Schindler}}, \bibinfo {author} {\bibfnamefont {S.~D.}\ \bibnamefont {Huber}},\ and\ \bibinfo {author} {\bibfnamefont {B.~A.}\ \bibnamefont {Bernevig}},\ }\bibfield  {title} {\bibinfo {title} {Superfluid weight bounds from symmetry and quantum geometry in flat bands},\ }\href {https://doi.org/10.1103/PhysRevLett.128.087002} {\bibfield  {journal} {\bibinfo  {journal} {Phys. Rev. Lett.}\ }\textbf {\bibinfo {volume} {128}},\ \bibinfo {pages} {087002} (\bibinfo {year} {2022})}\BibitemShut {NoStop}%
\bibitem [{\citenamefont {Hofmann}\ \emph {et~al.}(2023)\citenamefont {Hofmann}, \citenamefont {Berg},\ and\ \citenamefont {Chowdhury}}]{HofmannJohannesS2023}%
  \BibitemOpen
  \bibfield  {author} {\bibinfo {author} {\bibfnamefont {J.~S.}\ \bibnamefont {Hofmann}}, \bibinfo {author} {\bibfnamefont {E.}~\bibnamefont {Berg}},\ and\ \bibinfo {author} {\bibfnamefont {D.}~\bibnamefont {Chowdhury}},\ }\bibfield  {title} {\bibinfo {title} {Superconductivity, charge density wave, and supersolidity in flat bands with a tunable quantum metric},\ }\href {https://doi.org/10.1103/PhysRevLett.130.226001} {\bibfield  {journal} {\bibinfo  {journal} {Phys. Rev. Lett.}\ }\textbf {\bibinfo {volume} {130}},\ \bibinfo {pages} {226001} (\bibinfo {year} {2023})}\BibitemShut {NoStop}%
\bibitem [{\citenamefont {{Peotta}}\ \emph {et~al.}(2023)\citenamefont {{Peotta}}, \citenamefont {{Huhtinen}},\ and\ \citenamefont {{T{\"o}rm{\"a}}}}]{Peotta2023}%
  \BibitemOpen
  \bibfield  {author} {\bibinfo {author} {\bibfnamefont {S.}~\bibnamefont {{Peotta}}}, \bibinfo {author} {\bibfnamefont {K.-E.}\ \bibnamefont {{Huhtinen}}},\ and\ \bibinfo {author} {\bibfnamefont {P.}~\bibnamefont {{T{\"o}rm{\"a}}}},\ }\bibfield  {title} {\bibinfo {title} {{Quantum geometry in superfluidity and superconductivity}},\ }\href {https://arxiv.org/abs/2308.08248} {\bibfield  {journal} {\bibinfo  {journal} {arXiv:2308.08248}\ } (\bibinfo {year} {2023})}\BibitemShut {NoStop}%
\bibitem [{\citenamefont {{Sun}}\ \emph {et~al.}(2024)\citenamefont {{Sun}}, \citenamefont {{Yu}}, \citenamefont {{Chen}}, \citenamefont {{Hu}},\ and\ \citenamefont {{Law}}}]{2024sunziting}%
  \BibitemOpen
  \bibfield  {author} {\bibinfo {author} {\bibfnamefont {Z.-T.}\ \bibnamefont {{Sun}}}, \bibinfo {author} {\bibfnamefont {R.-P.}\ \bibnamefont {{Yu}}}, \bibinfo {author} {\bibfnamefont {S.~A.}\ \bibnamefont {{Chen}}}, \bibinfo {author} {\bibfnamefont {J.-X.}\ \bibnamefont {{Hu}}},\ and\ \bibinfo {author} {\bibfnamefont {K.~T.}\ \bibnamefont {{Law}}},\ }\bibfield  {title} {\bibinfo {title} {{Flat-band FFLO State from Quantum Geometric Discrepancy}},\ }\href {https://arxiv.org/abs/2408.00548} {\bibfield  {journal} {\bibinfo  {journal} {arXiv:2408.00548}\ } (\bibinfo {year} {2024})}\BibitemShut {NoStop}%
\bibitem [{\citenamefont {Roy}(2014)}]{RoyRahul2014}%
  \BibitemOpen
  \bibfield  {author} {\bibinfo {author} {\bibfnamefont {R.}~\bibnamefont {Roy}},\ }\bibfield  {title} {\bibinfo {title} {Band geometry of fractional topological insulators},\ }\href {https://doi.org/10.1103/PhysRevB.90.165139} {\bibfield  {journal} {\bibinfo  {journal} {Phys. Rev. B}\ }\textbf {\bibinfo {volume} {90}},\ \bibinfo {pages} {165139} (\bibinfo {year} {2014})}\BibitemShut {NoStop}%
\bibitem [{\citenamefont {Ozawa}\ and\ \citenamefont {Mera}(2021)}]{OzawaTomoki2021}%
  \BibitemOpen
  \bibfield  {author} {\bibinfo {author} {\bibfnamefont {T.}~\bibnamefont {Ozawa}}\ and\ \bibinfo {author} {\bibfnamefont {B.}~\bibnamefont {Mera}},\ }\bibfield  {title} {\bibinfo {title} {Relations between topology and the quantum metric for chern insulators},\ }\href {https://doi.org/10.1103/PhysRevB.104.045103} {\bibfield  {journal} {\bibinfo  {journal} {Phys. Rev. B}\ }\textbf {\bibinfo {volume} {104}},\ \bibinfo {pages} {045103} (\bibinfo {year} {2021})}\BibitemShut {NoStop}%
\bibitem [{\citenamefont {Okuma}(2024)}]{OkumaNobuyuki2024}%
  \BibitemOpen
  \bibfield  {author} {\bibinfo {author} {\bibfnamefont {N.}~\bibnamefont {Okuma}},\ }\bibfield  {title} {\bibinfo {title} {Constructing vortex functions and basis states of chern insulators: Ideal condition, inequality from index theorem, and coherentlike states on the von neumann lattice},\ }\href {https://doi.org/10.1103/PhysRevB.110.245112} {\bibfield  {journal} {\bibinfo  {journal} {Phys. Rev. B}\ }\textbf {\bibinfo {volume} {110}},\ \bibinfo {pages} {245112} (\bibinfo {year} {2024})}\BibitemShut {NoStop}%
\bibitem [{\citenamefont {Kwon}\ and\ \citenamefont {Yang}(2024{\natexlab{a}})}]{KwonSoonhyun2024}%
  \BibitemOpen
  \bibfield  {author} {\bibinfo {author} {\bibfnamefont {S.}~\bibnamefont {Kwon}}\ and\ \bibinfo {author} {\bibfnamefont {B.-J.}\ \bibnamefont {Yang}},\ }\bibfield  {title} {\bibinfo {title} {Quantum geometric bound and ideal condition for euler band topology},\ }\href {https://doi.org/10.1103/PhysRevB.109.L161111} {\bibfield  {journal} {\bibinfo  {journal} {Phys. Rev. B}\ }\textbf {\bibinfo {volume} {109}},\ \bibinfo {pages} {L161111} (\bibinfo {year} {2024}{\natexlab{a}})}\BibitemShut {NoStop}%
\bibitem [{\citenamefont {Ding}\ \emph {et~al.}(2024)\citenamefont {Ding}, \citenamefont {Zhang}, \citenamefont {Liu}, \citenamefont {Wang}, \citenamefont {Zhang},\ and\ \citenamefont {Zhu}}]{DingHaiTao2024}%
  \BibitemOpen
  \bibfield  {author} {\bibinfo {author} {\bibfnamefont {H.-T.}\ \bibnamefont {Ding}}, \bibinfo {author} {\bibfnamefont {C.-X.}\ \bibnamefont {Zhang}}, \bibinfo {author} {\bibfnamefont {J.-X.}\ \bibnamefont {Liu}}, \bibinfo {author} {\bibfnamefont {J.-T.}\ \bibnamefont {Wang}}, \bibinfo {author} {\bibfnamefont {D.-W.}\ \bibnamefont {Zhang}},\ and\ \bibinfo {author} {\bibfnamefont {S.-L.}\ \bibnamefont {Zhu}},\ }\bibfield  {title} {\bibinfo {title} {Non-abelian quantum geometric tensor in degenerate topological semimetals},\ }\href {https://doi.org/10.1103/PhysRevA.109.043305} {\bibfield  {journal} {\bibinfo  {journal} {Phys. Rev. A}\ }\textbf {\bibinfo {volume} {109}},\ \bibinfo {pages} {043305} (\bibinfo {year} {2024})}\BibitemShut {NoStop}%
\bibitem [{\citenamefont {Romeral}\ \emph {et~al.}(2025)\citenamefont {Romeral}, \citenamefont {Cummings},\ and\ \citenamefont {Roche}}]{Romeral2024}%
  \BibitemOpen
  \bibfield  {author} {\bibinfo {author} {\bibfnamefont {J.~M.}\ \bibnamefont {Romeral}}, \bibinfo {author} {\bibfnamefont {A.~W.}\ \bibnamefont {Cummings}},\ and\ \bibinfo {author} {\bibfnamefont {S.}~\bibnamefont {Roche}},\ }\bibfield  {title} {\bibinfo {title} {Scaling of the integrated quantum metric in disordered topological phases},\ }\href {https://doi.org/10.1103/PhysRevB.111.134201} {\bibfield  {journal} {\bibinfo  {journal} {Phys. Rev. B}\ }\textbf {\bibinfo {volume} {111}},\ \bibinfo {pages} {134201} (\bibinfo {year} {2025})}\BibitemShut {NoStop}%
\bibitem [{\citenamefont {Yu}\ \emph {et~al.}(2025)\citenamefont {Yu}, \citenamefont {Herzog-Arbeitman},\ and\ \citenamefont {Bernevig}}]{Yu2025}%
  \BibitemOpen
  \bibfield  {author} {\bibinfo {author} {\bibfnamefont {J.}~\bibnamefont {Yu}}, \bibinfo {author} {\bibfnamefont {J.}~\bibnamefont {Herzog-Arbeitman}},\ and\ \bibinfo {author} {\bibfnamefont {B.~A.}\ \bibnamefont {Bernevig}},\ }\bibfield  {title} {\bibinfo {title} {Universal wilson loop bound of quantum geometry},\ }\href {https://doi.org/10.1103/mp2c-zzkt} {\bibfield  {journal} {\bibinfo  {journal} {Phys. Rev. Lett.}\ }\textbf {\bibinfo {volume} {135}},\ \bibinfo {pages} {086401} (\bibinfo {year} {2025})}\BibitemShut {NoStop}%
\bibitem [{\citenamefont {{Chiu}}(2025)}]{chiupokman2025}%
  \BibitemOpen
  \bibfield  {author} {\bibinfo {author} {\bibfnamefont {P.~M.}\ \bibnamefont {{Chiu}}},\ }\bibfield  {title} {\bibinfo {title} {{Topological Signatures of the Optical Bound on Maximal Berry Curvature: Applications to Two-Dimensional Time-Reversal-Symmetric Insulators}},\ }\href {https://arxiv.org/abs/2501.17671} {\bibfield  {journal} {\bibinfo  {journal} {arXiv:2501.17671}\ } (\bibinfo {year} {2025})}\BibitemShut {NoStop}%
\bibitem [{\citenamefont {{Jankowski}}\ \emph {et~al.}(2025)\citenamefont {{Jankowski}}, \citenamefont {{Slager}},\ and\ \citenamefont {{Lange}}}]{Jankowski2025}%
  \BibitemOpen
  \bibfield  {author} {\bibinfo {author} {\bibfnamefont {W.~J.}\ \bibnamefont {{Jankowski}}}, \bibinfo {author} {\bibfnamefont {R.-J.}\ \bibnamefont {{Slager}}},\ and\ \bibinfo {author} {\bibfnamefont {G.~F.}\ \bibnamefont {{Lange}}},\ }\bibfield  {title} {\bibinfo {title} {{Quantum geometric bounds in spinful systems with trivial band topology}},\ }\href {https://arxiv.org/abs/2501.16428} {\bibfield  {journal} {\bibinfo  {journal} {arXiv:2501.16428}\ } (\bibinfo {year} {2025})}\BibitemShut {NoStop}%
\bibitem [{\citenamefont {Jankowski}\ \emph {et~al.}(2025)\citenamefont {Jankowski}, \citenamefont {Morris}, \citenamefont {Bouhon}, \citenamefont {\"Unal},\ and\ \citenamefont {Slager}}]{Jankowski2025a}%
  \BibitemOpen
  \bibfield  {author} {\bibinfo {author} {\bibfnamefont {W.~J.}\ \bibnamefont {Jankowski}}, \bibinfo {author} {\bibfnamefont {A.~S.}\ \bibnamefont {Morris}}, \bibinfo {author} {\bibfnamefont {A.}~\bibnamefont {Bouhon}}, \bibinfo {author} {\bibfnamefont {F.~N.}\ \bibnamefont {\"Unal}},\ and\ \bibinfo {author} {\bibfnamefont {R.-J.}\ \bibnamefont {Slager}},\ }\bibfield  {title} {\bibinfo {title} {Optical manifestations and bounds of topological euler class},\ }\href {https://doi.org/10.1103/PhysRevB.111.L081103} {\bibfield  {journal} {\bibinfo  {journal} {Phys. Rev. B}\ }\textbf {\bibinfo {volume} {111}},\ \bibinfo {pages} {L081103} (\bibinfo {year} {2025})}\BibitemShut {NoStop}%
\bibitem [{\citenamefont {Mitscherling}\ and\ \citenamefont {Holder}(2022)}]{Mitscherling2022}%
  \BibitemOpen
  \bibfield  {author} {\bibinfo {author} {\bibfnamefont {J.}~\bibnamefont {Mitscherling}}\ and\ \bibinfo {author} {\bibfnamefont {T.}~\bibnamefont {Holder}},\ }\bibfield  {title} {\bibinfo {title} {Bound on resistivity in flat-band materials due to the quantum metric},\ }\href {https://doi.org/10.1103/PhysRevB.105.085154} {\bibfield  {journal} {\bibinfo  {journal} {Phys. Rev. B}\ }\textbf {\bibinfo {volume} {105}},\ \bibinfo {pages} {085154} (\bibinfo {year} {2022})}\BibitemShut {NoStop}%
\bibitem [{\citenamefont {{Xiao}}\ and\ \citenamefont {{Hao}}(2024)}]{Yuhang2024}%
  \BibitemOpen
  \bibfield  {author} {\bibinfo {author} {\bibfnamefont {Y.}~\bibnamefont {{Xiao}}}\ and\ \bibinfo {author} {\bibfnamefont {N.}~\bibnamefont {{Hao}}},\ }\bibfield  {title} {\bibinfo {title} {{Quantum Geometric Effects on the Higgs Mode in Flat-band Superconductors}},\ }\href {https://arxiv.org/abs/2409.10891} {\bibfield  {journal} {\bibinfo  {journal} {arXiv:2409.10891}\ } (\bibinfo {year} {2024})}\BibitemShut {NoStop}%
\bibitem [{\citenamefont {{Ying}}\ and\ \citenamefont {{Law}}(2024)}]{Xuzhe2024arXiv}%
  \BibitemOpen
  \bibfield  {author} {\bibinfo {author} {\bibfnamefont {X.}~\bibnamefont {{Ying}}}\ and\ \bibinfo {author} {\bibfnamefont {K.~T.}\ \bibnamefont {{Law}}},\ }\bibfield  {title} {\bibinfo {title} {{Flat band excitons and quantum metric}},\ }\href {https://arxiv.org/abs/2407.00325} {\bibfield  {journal} {\bibinfo  {journal} {arXiv:2407.00325}\ } (\bibinfo {year} {2024})}\BibitemShut {NoStop}%
\bibitem [{\citenamefont {Ying}\ and\ \citenamefont {Li}(2025)}]{Xuzhe2024}%
  \BibitemOpen
  \bibfield  {author} {\bibinfo {author} {\bibfnamefont {X.}~\bibnamefont {Ying}}\ and\ \bibinfo {author} {\bibfnamefont {K.}~\bibnamefont {Li}},\ }\bibfield  {title} {\bibinfo {title} {Quantum metric driven transition between superfluid and incoherent fluid},\ }\href {https://doi.org/10.1103/3pvf-1ckd} {\bibfield  {journal} {\bibinfo  {journal} {Phys. Rev. B}\ }\textbf {\bibinfo {volume} {112}},\ \bibinfo {pages} {014518} (\bibinfo {year} {2025})}\BibitemShut {NoStop}%
\bibitem [{\citenamefont {{Chau}}\ \emph {et~al.}(2024)\citenamefont {{Chau}}, \citenamefont {{Xiang}}, \citenamefont {{Chen}},\ and\ \citenamefont {{Law}}}]{ChunWang2024}%
  \BibitemOpen
  \bibfield  {author} {\bibinfo {author} {\bibfnamefont {C.~W.}\ \bibnamefont {{Chau}}}, \bibinfo {author} {\bibfnamefont {T.}~\bibnamefont {{Xiang}}}, \bibinfo {author} {\bibfnamefont {S.~A.}\ \bibnamefont {{Chen}}},\ and\ \bibinfo {author} {\bibfnamefont {K.~T.}\ \bibnamefont {{Law}}},\ }\bibfield  {title} {\bibinfo {title} {{Disorder-induced delocalization in flat-band systems with quantum geometry}},\ }\href {https://arxiv.org/abs/2412.19056} {\bibfield  {journal} {\bibinfo  {journal} {arXiv:2412.19056}\ } (\bibinfo {year} {2024})}\BibitemShut {NoStop}%
\bibitem [{\citenamefont {Wen}\ \emph {et~al.}(2025)\citenamefont {Wen}, \citenamefont {Xie}, \citenamefont {Auerbach},\ and\ \citenamefont {Uchoa}}]{Kevin2025}%
  \BibitemOpen
  \bibfield  {author} {\bibinfo {author} {\bibfnamefont {K.}~\bibnamefont {Wen}}, \bibinfo {author} {\bibfnamefont {H.-Y.}\ \bibnamefont {Xie}}, \bibinfo {author} {\bibfnamefont {A.}~\bibnamefont {Auerbach}},\ and\ \bibinfo {author} {\bibfnamefont {B.}~\bibnamefont {Uchoa}},\ }\bibfield  {title} {\bibinfo {title} {Thermal and thermoelectric transport in flat bands with nontrivial quantum geometry},\ }\href {https://doi.org/10.1103/PhysRevB.111.205140} {\bibfield  {journal} {\bibinfo  {journal} {Phys. Rev. B}\ }\textbf {\bibinfo {volume} {111}},\ \bibinfo {pages} {205140} (\bibinfo {year} {2025})}\BibitemShut {NoStop}%
\bibitem [{\citenamefont {Chen}\ and\ \citenamefont {Law}(2024)}]{ChenShuaiA2024}%
  \BibitemOpen
  \bibfield  {author} {\bibinfo {author} {\bibfnamefont {S.~A.}\ \bibnamefont {Chen}}\ and\ \bibinfo {author} {\bibfnamefont {K.~T.}\ \bibnamefont {Law}},\ }\bibfield  {title} {\bibinfo {title} {Ginzburg-landau theory of flat-band superconductors with quantum metric},\ }\href {https://doi.org/10.1103/PhysRevLett.132.026002} {\bibfield  {journal} {\bibinfo  {journal} {Phys. Rev. Lett.}\ }\textbf {\bibinfo {volume} {132}},\ \bibinfo {pages} {026002} (\bibinfo {year} {2024})}\BibitemShut {NoStop}%
\bibitem [{\citenamefont {Li}\ \emph {et~al.}(2025)\citenamefont {Li}, \citenamefont {Deng}, \citenamefont {Chen}, \citenamefont {Efetov},\ and\ \citenamefont {Law}}]{lizhong2024}%
  \BibitemOpen
  \bibfield  {author} {\bibinfo {author} {\bibfnamefont {Z.~C.~F.}\ \bibnamefont {Li}}, \bibinfo {author} {\bibfnamefont {Y.}~\bibnamefont {Deng}}, \bibinfo {author} {\bibfnamefont {S.~A.}\ \bibnamefont {Chen}}, \bibinfo {author} {\bibfnamefont {D.~K.}\ \bibnamefont {Efetov}},\ and\ \bibinfo {author} {\bibfnamefont {K.~T.}\ \bibnamefont {Law}},\ }\bibfield  {title} {\bibinfo {title} {Flat band josephson junctions with quantum metric},\ }\href {https://doi.org/10.1103/PhysRevResearch.7.023273} {\bibfield  {journal} {\bibinfo  {journal} {Phys. Rev. Res.}\ }\textbf {\bibinfo {volume} {7}},\ \bibinfo {pages} {023273} (\bibinfo {year} {2025})}\BibitemShut {NoStop}%
\bibitem [{\citenamefont {Hu}\ \emph {et~al.}(2025)\citenamefont {Hu}, \citenamefont {Chen},\ and\ \citenamefont {Law}}]{Hu2025}%
  \BibitemOpen
  \bibfield  {author} {\bibinfo {author} {\bibfnamefont {J.-X.}\ \bibnamefont {Hu}}, \bibinfo {author} {\bibfnamefont {S.~A.}\ \bibnamefont {Chen}},\ and\ \bibinfo {author} {\bibfnamefont {K.~T.}\ \bibnamefont {Law}},\ }\bibfield  {title} {\bibinfo {title} {Anomalous coherence length in superconductors with quantum metric},\ }\href {https://doi.org/10.1038/s42005-024-01930-0} {\bibfield  {journal} {\bibinfo  {journal} {Communications Physics}\ }\textbf {\bibinfo {volume} {8}},\ \bibinfo {pages} {20} (\bibinfo {year} {2025})}\BibitemShut {NoStop}%
\bibitem [{\citenamefont {{Ma}}\ \emph {et~al.}(2025)\citenamefont {{Ma}}, \citenamefont {{Hu}},\ and\ \citenamefont {{Law}}}]{2025arXivxinglei}%
  \BibitemOpen
  \bibfield  {author} {\bibinfo {author} {\bibfnamefont {X.-L.}\ \bibnamefont {{Ma}}}, \bibinfo {author} {\bibfnamefont {J.-X.}\ \bibnamefont {{Hu}}},\ and\ \bibinfo {author} {\bibfnamefont {K.~T.}\ \bibnamefont {{Law}}},\ }\bibfield  {title} {\bibinfo {title} {{Universal Boundary-Modes Localization from Quantum Metric Length}},\ }\href {https://arxiv.org/abs/2509.05114} {\bibfield  {journal} {\bibinfo  {journal} {arXiv:2509.05114}\ } (\bibinfo {year} {2025})}\BibitemShut {NoStop}%
\bibitem [{\citenamefont {Guo}\ \emph {et~al.}(2025)\citenamefont {Guo}, \citenamefont {Ma}, \citenamefont {Ying},\ and\ \citenamefont {Law}}]{xingyao2024}%
  \BibitemOpen
  \bibfield  {author} {\bibinfo {author} {\bibfnamefont {X.}~\bibnamefont {Guo}}, \bibinfo {author} {\bibfnamefont {X.}~\bibnamefont {Ma}}, \bibinfo {author} {\bibfnamefont {X.}~\bibnamefont {Ying}},\ and\ \bibinfo {author} {\bibfnamefont {K.~T.}\ \bibnamefont {Law}},\ }\bibfield  {title} {\bibinfo {title} {Majorana zero modes in the lieb-kitaev model with tunable quantum metric},\ }\href {https://doi.org/10.1103/l3c7-knqm} {\bibfield  {journal} {\bibinfo  {journal} {Phys. Rev. Lett.}\ }\textbf {\bibinfo {volume} {135}},\ \bibinfo {pages} {076601} (\bibinfo {year} {2025})}\BibitemShut {NoStop}%
\bibitem [{\citenamefont {Ryu}\ \emph {et~al.}(2010)\citenamefont {Ryu}, \citenamefont {Schnyder}, \citenamefont {Furusaki},\ and\ \citenamefont {Ludwig}}]{Ryu2010}%
  \BibitemOpen
  \bibfield  {author} {\bibinfo {author} {\bibfnamefont {S.}~\bibnamefont {Ryu}}, \bibinfo {author} {\bibfnamefont {A.~P.}\ \bibnamefont {Schnyder}}, \bibinfo {author} {\bibfnamefont {A.}~\bibnamefont {Furusaki}},\ and\ \bibinfo {author} {\bibfnamefont {A.~W.~W.}\ \bibnamefont {Ludwig}},\ }\bibfield  {title} {\bibinfo {title} {Topological insulators and superconductors: tenfold way and dimensional hierarchy},\ }\href {https://doi.org/10.1088/1367-2630/12/6/065010} {\bibfield  {journal} {\bibinfo  {journal} {New Journal of Physics}\ }\textbf {\bibinfo {volume} {12}},\ \bibinfo {pages} {065010} (\bibinfo {year} {2010})}\BibitemShut {NoStop}%
\bibitem [{\citenamefont {Fu}\ and\ \citenamefont {Kane}(2007)}]{FuLiang2027inversion}%
  \BibitemOpen
  \bibfield  {author} {\bibinfo {author} {\bibfnamefont {L.}~\bibnamefont {Fu}}\ and\ \bibinfo {author} {\bibfnamefont {C.~L.}\ \bibnamefont {Kane}},\ }\bibfield  {title} {\bibinfo {title} {Topological insulators with inversion symmetry},\ }\href {https://doi.org/10.1103/PhysRevB.76.045302} {\bibfield  {journal} {\bibinfo  {journal} {Phys. Rev. B}\ }\textbf {\bibinfo {volume} {76}},\ \bibinfo {pages} {045302} (\bibinfo {year} {2007})}\BibitemShut {NoStop}%
\bibitem [{\citenamefont {Fu}\ \emph {et~al.}(2007{\natexlab{b}})\citenamefont {Fu}, \citenamefont {Kane},\ and\ \citenamefont {Mele}}]{Fu200C}%
  \BibitemOpen
  \bibfield  {author} {\bibinfo {author} {\bibfnamefont {L.}~\bibnamefont {Fu}}, \bibinfo {author} {\bibfnamefont {C.~L.}\ \bibnamefont {Kane}},\ and\ \bibinfo {author} {\bibfnamefont {E.~J.}\ \bibnamefont {Mele}},\ }\bibfield  {title} {\bibinfo {title} {Topological insulators in three dimensions},\ }\href {https://doi.org/10.1103/PhysRevLett.98.106803} {\bibfield  {journal} {\bibinfo  {journal} {Phys. Rev. Lett.}\ }\textbf {\bibinfo {volume} {98}},\ \bibinfo {pages} {106803} (\bibinfo {year} {2007}{\natexlab{b}})}\BibitemShut {NoStop}%
\bibitem [{\citenamefont {Luo}\ and\ \citenamefont {Wu}(2025)}]{2024BTIluo}%
  \BibitemOpen
  \bibfield  {author} {\bibinfo {author} {\bibfnamefont {X.-J.}\ \bibnamefont {Luo}}\ and\ \bibinfo {author} {\bibfnamefont {F.}~\bibnamefont {Wu}},\ }\bibfield  {title} {\bibinfo {title} {Boundary topological insulators and superconductors of altland-zirnbauer tenfold classes},\ }\href {https://doi.org/10.1103/xfkg-drhb} {\bibfield  {journal} {\bibinfo  {journal} {Phys. Rev. B}\ }\textbf {\bibinfo {volume} {111}},\ \bibinfo {pages} {245143} (\bibinfo {year} {2025})}\BibitemShut {NoStop}%
\bibitem [{\citenamefont {Su}\ \emph {et~al.}(1979)\citenamefont {Su}, \citenamefont {Schrieffer},\ and\ \citenamefont {Heeger}}]{Su1979}%
  \BibitemOpen
  \bibfield  {author} {\bibinfo {author} {\bibfnamefont {W.~P.}\ \bibnamefont {Su}}, \bibinfo {author} {\bibfnamefont {J.~R.}\ \bibnamefont {Schrieffer}},\ and\ \bibinfo {author} {\bibfnamefont {A.~J.}\ \bibnamefont {Heeger}},\ }\bibfield  {title} {\bibinfo {title} {Solitons in polyacetylene},\ }\href {https://doi.org/10.1103/PhysRevLett.42.1698} {\bibfield  {journal} {\bibinfo  {journal} {Phys. Rev. Lett.}\ }\textbf {\bibinfo {volume} {42}},\ \bibinfo {pages} {1698} (\bibinfo {year} {1979})}\BibitemShut {NoStop}%
\bibitem [{\citenamefont {Qi}\ \emph {et~al.}(2006)\citenamefont {Qi}, \citenamefont {Wu},\ and\ \citenamefont {Zhang}}]{Qi2006}%
  \BibitemOpen
  \bibfield  {author} {\bibinfo {author} {\bibfnamefont {X.-L.}\ \bibnamefont {Qi}}, \bibinfo {author} {\bibfnamefont {Y.-S.}\ \bibnamefont {Wu}},\ and\ \bibinfo {author} {\bibfnamefont {S.-C.}\ \bibnamefont {Zhang}},\ }\bibfield  {title} {\bibinfo {title} {Topological quantization of the spin hall effect in two-dimensional paramagnetic semiconductors},\ }\href {https://doi.org/10.1103/PhysRevB.74.085308} {\bibfield  {journal} {\bibinfo  {journal} {Phys. Rev. B}\ }\textbf {\bibinfo {volume} {74}},\ \bibinfo {pages} {085308} (\bibinfo {year} {2006})}\BibitemShut {NoStop}%
\bibitem [{\citenamefont {Mera}\ \emph {et~al.}(2022)\citenamefont {Mera}, \citenamefont {Zhang},\ and\ \citenamefont {Goldman}}]{Mera2022}%
  \BibitemOpen
  \bibfield  {author} {\bibinfo {author} {\bibfnamefont {B.}~\bibnamefont {Mera}}, \bibinfo {author} {\bibfnamefont {A.}~\bibnamefont {Zhang}},\ and\ \bibinfo {author} {\bibfnamefont {N.}~\bibnamefont {Goldman}},\ }\bibfield  {title} {\bibinfo {title} {Relating the topology of dirac hamiltonians to quantum geometry: When the quantum metric dictates chern numbers and winding numbers},\ }\href {https://scipost.org/SciPostPhys.12.1.018} {\bibfield  {journal} {\bibinfo  {journal} {SciPost Physics}\ }\textbf {\bibinfo {volume} {12}},\ \bibinfo {pages} {018} (\bibinfo {year} {2022})}\BibitemShut {NoStop}%
\bibitem [{\citenamefont {Kwon}\ and\ \citenamefont {Yang}(2024{\natexlab{b}})}]{Kwon2024}%
  \BibitemOpen
  \bibfield  {author} {\bibinfo {author} {\bibfnamefont {S.}~\bibnamefont {Kwon}}\ and\ \bibinfo {author} {\bibfnamefont {B.-J.}\ \bibnamefont {Yang}},\ }\bibfield  {title} {\bibinfo {title} {Quantum geometric bound and ideal condition for euler band topology},\ }\href {https://doi.org/10.1103/PhysRevB.109.L161111} {\bibfield  {journal} {\bibinfo  {journal} {Phys. Rev. B}\ }\textbf {\bibinfo {volume} {109}},\ \bibinfo {pages} {L161111} (\bibinfo {year} {2024}{\natexlab{b}})}\BibitemShut {NoStop}%
\bibitem [{\citenamefont {{Onishi}}\ and\ \citenamefont {{Fu}}(2024{\natexlab{a}})}]{Onishi2024a}%
  \BibitemOpen
  \bibfield  {author} {\bibinfo {author} {\bibfnamefont {Y.}~\bibnamefont {{Onishi}}}\ and\ \bibinfo {author} {\bibfnamefont {L.}~\bibnamefont {{Fu}}},\ }\bibfield  {title} {\bibinfo {title} {{Quantum weight}},\ }\href {https://arxiv.org/abs/2401.13847} {\bibfield  {journal} {\bibinfo  {journal} {arXiv:2401.13847}\ } (\bibinfo {year} {2024}{\natexlab{a}})}\BibitemShut {NoStop}%
\bibitem [{\citenamefont {{Onishi}}\ and\ \citenamefont {{Fu}}(2024{\natexlab{b}})}]{Onishi2024b}%
  \BibitemOpen
  \bibfield  {author} {\bibinfo {author} {\bibfnamefont {Y.}~\bibnamefont {{Onishi}}}\ and\ \bibinfo {author} {\bibfnamefont {L.}~\bibnamefont {{Fu}}},\ }\bibfield  {title} {\bibinfo {title} {{Quantum weight: A fundamental property of quantum many-body systems}},\ }\href {https://arxiv.org/abs/2406.06783} {\bibfield  {journal} {\bibinfo  {journal} {arXiv:2406.06783}\ } (\bibinfo {year} {2024}{\natexlab{b}})}\BibitemShut {NoStop}%
\bibitem [{\citenamefont {Asb'oth}\ \emph {et~al.}(2015)\citenamefont {Asb'oth}, \citenamefont {Oroszl'any},\ and\ \citenamefont {P'alyi}}]{Asboth2015ASC}%
  \BibitemOpen
  \bibfield  {author} {\bibinfo {author} {\bibfnamefont {J.~K.}\ \bibnamefont {Asb'oth}}, \bibinfo {author} {\bibfnamefont {L.}~\bibnamefont {Oroszl'any}},\ and\ \bibinfo {author} {\bibfnamefont {A.}~\bibnamefont {P'alyi}},\ }\bibfield  {title} {\bibinfo {title} {A short course on topological insulators: Band-structure topology and edge states in one and two dimensions},\ }\href {https://api.semanticscholar.org/CorpusID:119271692} {\bibfield  {journal} {\bibinfo  {journal} {arXiv: Mesoscale and Nanoscale Physics}\ }\textbf {\bibinfo {volume} {919}} (\bibinfo {year} {2015})}\BibitemShut {NoStop}%
\bibitem [{\citenamefont {Zhang}\ \emph {et~al.}(2021)\citenamefont {Zhang}, \citenamefont {Devakul},\ and\ \citenamefont {Fu}}]{Zhang2021}%
  \BibitemOpen
  \bibfield  {author} {\bibinfo {author} {\bibfnamefont {Y.}~\bibnamefont {Zhang}}, \bibinfo {author} {\bibfnamefont {T.}~\bibnamefont {Devakul}},\ and\ \bibinfo {author} {\bibfnamefont {L.}~\bibnamefont {Fu}},\ }\bibfield  {title} {\bibinfo {title} {{Spin-textured Chern bands in AB-stacked transition metal dichalcogenide bilayers}},\ }\href {https://doi.org/10.1073/pnas.2112673118} {\bibfield  {journal} {\bibinfo  {journal} {Proc. Natl. Acad. Sci. U.S.A.}\ }\textbf {\bibinfo {volume} {118}},\ \bibinfo {pages} {e2112673118} (\bibinfo {year} {2021})}\BibitemShut {NoStop}%
\bibitem [{\citenamefont {Luo}\ \emph {et~al.}(2023{\natexlab{b}})\citenamefont {Luo}, \citenamefont {Wang},\ and\ \citenamefont {Wu}}]{Symmetricluo}%
  \BibitemOpen
  \bibfield  {author} {\bibinfo {author} {\bibfnamefont {X.-J.}\ \bibnamefont {Luo}}, \bibinfo {author} {\bibfnamefont {M.}~\bibnamefont {Wang}},\ and\ \bibinfo {author} {\bibfnamefont {F.}~\bibnamefont {Wu}},\ }\bibfield  {title} {\bibinfo {title} {{Symmetric Wannier states and tight-binding model for quantum spin Hall bands in $AB$-stacked ${\text{MoTe}}_{2}/{\text{WSe}}_{2}$}},\ }\href {https://doi.org/10.1103/PhysRevB.107.235127} {\bibfield  {journal} {\bibinfo  {journal} {Phys. Rev. B}\ }\textbf {\bibinfo {volume} {107}},\ \bibinfo {pages} {235127} (\bibinfo {year} {2023}{\natexlab{b}})}\BibitemShut {NoStop}%
\bibitem [{\citenamefont {Dzero}\ \emph {et~al.}(2010)\citenamefont {Dzero}, \citenamefont {Sun}, \citenamefont {Galitski},\ and\ \citenamefont {Coleman}}]{PhysRevLett.104.106408}%
  \BibitemOpen
  \bibfield  {author} {\bibinfo {author} {\bibfnamefont {M.}~\bibnamefont {Dzero}}, \bibinfo {author} {\bibfnamefont {K.}~\bibnamefont {Sun}}, \bibinfo {author} {\bibfnamefont {V.}~\bibnamefont {Galitski}},\ and\ \bibinfo {author} {\bibfnamefont {P.}~\bibnamefont {Coleman}},\ }\bibfield  {title} {\bibinfo {title} {Topological kondo insulators},\ }\href {https://doi.org/10.1103/PhysRevLett.104.106408} {\bibfield  {journal} {\bibinfo  {journal} {Phys. Rev. Lett.}\ }\textbf {\bibinfo {volume} {104}},\ \bibinfo {pages} {106408} (\bibinfo {year} {2010})}\BibitemShut {NoStop}%
\bibitem [{\citenamefont {Neupane}\ \emph {et~al.}(2013)\citenamefont {Neupane}, \citenamefont {Alidoust}, \citenamefont {Xu}, \citenamefont {Kondo}, \citenamefont {Ishida}, \citenamefont {Kim}, \citenamefont {Liu}, \citenamefont {Belopolski}, \citenamefont {Jo}, \citenamefont {Chang}, \citenamefont {Jeng}, \citenamefont {Durakiewicz}, \citenamefont {Balicas}, \citenamefont {Lin}, \citenamefont {Bansil}, \citenamefont {Shin}, \citenamefont {Fisk},\ and\ \citenamefont {Hasan}}]{Neupane2013}%
  \BibitemOpen
  \bibfield  {author} {\bibinfo {author} {\bibfnamefont {M.}~\bibnamefont {Neupane}}, \bibinfo {author} {\bibfnamefont {N.}~\bibnamefont {Alidoust}}, \bibinfo {author} {\bibfnamefont {S.-Y.}\ \bibnamefont {Xu}}, \bibinfo {author} {\bibfnamefont {T.}~\bibnamefont {Kondo}}, \bibinfo {author} {\bibfnamefont {Y.}~\bibnamefont {Ishida}}, \bibinfo {author} {\bibfnamefont {D.~J.}\ \bibnamefont {Kim}}, \bibinfo {author} {\bibfnamefont {C.}~\bibnamefont {Liu}}, \bibinfo {author} {\bibfnamefont {I.}~\bibnamefont {Belopolski}}, \bibinfo {author} {\bibfnamefont {Y.~J.}\ \bibnamefont {Jo}}, \bibinfo {author} {\bibfnamefont {T.-R.}\ \bibnamefont {Chang}}, \bibinfo {author} {\bibfnamefont {H.-T.}\ \bibnamefont {Jeng}}, \bibinfo {author} {\bibfnamefont {T.}~\bibnamefont {Durakiewicz}}, \bibinfo {author} {\bibfnamefont {L.}~\bibnamefont {Balicas}}, \bibinfo {author} {\bibfnamefont {H.}~\bibnamefont {Lin}}, \bibinfo {author} {\bibfnamefont {A.}~\bibnamefont {Bansil}}, \bibinfo {author} {\bibfnamefont {S.}~\bibnamefont {Shin}},
  \bibinfo {author} {\bibfnamefont {Z.}~\bibnamefont {Fisk}},\ and\ \bibinfo {author} {\bibfnamefont {M.~Z.}\ \bibnamefont {Hasan}},\ }\bibfield  {title} {\bibinfo {title} {Surface electronic structure of the topological kondo-insulator candidate correlated electron system smb6},\ }\href {https://doi.org/10.1038/ncomms3991} {\bibfield  {journal} {\bibinfo  {journal} {Nature Communications}\ }\textbf {\bibinfo {volume} {4}},\ \bibinfo {pages} {2991} (\bibinfo {year} {2013})}\BibitemShut {NoStop}%
\bibitem [{\citenamefont {Xue}\ \emph {et~al.}(2022)\citenamefont {Xue}, \citenamefont {Yang},\ and\ \citenamefont {Zhang}}]{Xue2022}%
  \BibitemOpen
  \bibfield  {author} {\bibinfo {author} {\bibfnamefont {H.}~\bibnamefont {Xue}}, \bibinfo {author} {\bibfnamefont {Y.}~\bibnamefont {Yang}},\ and\ \bibinfo {author} {\bibfnamefont {B.}~\bibnamefont {Zhang}},\ }\bibfield  {title} {\bibinfo {title} {Topological acoustics},\ }\href {https://doi.org/10.1038/s41578-022-00465-6} {\bibfield  {journal} {\bibinfo  {journal} {Nature Reviews Materials}\ }\textbf {\bibinfo {volume} {7}},\ \bibinfo {pages} {974} (\bibinfo {year} {2022})}\BibitemShut {NoStop}%
\bibitem [{\citenamefont {Ozawa}\ \emph {et~al.}(2019)\citenamefont {Ozawa}, \citenamefont {Price}, \citenamefont {Amo}, \citenamefont {Goldman}, \citenamefont {Hafezi}, \citenamefont {Lu}, \citenamefont {Rechtsman}, \citenamefont {Schuster}, \citenamefont {Simon}, \citenamefont {Zilberberg},\ and\ \citenamefont {Carusotto}}]{OzawaTomoki2019}%
  \BibitemOpen
  \bibfield  {author} {\bibinfo {author} {\bibfnamefont {T.}~\bibnamefont {Ozawa}}, \bibinfo {author} {\bibfnamefont {H.~M.}\ \bibnamefont {Price}}, \bibinfo {author} {\bibfnamefont {A.}~\bibnamefont {Amo}}, \bibinfo {author} {\bibfnamefont {N.}~\bibnamefont {Goldman}}, \bibinfo {author} {\bibfnamefont {M.}~\bibnamefont {Hafezi}}, \bibinfo {author} {\bibfnamefont {L.}~\bibnamefont {Lu}}, \bibinfo {author} {\bibfnamefont {M.~C.}\ \bibnamefont {Rechtsman}}, \bibinfo {author} {\bibfnamefont {D.}~\bibnamefont {Schuster}}, \bibinfo {author} {\bibfnamefont {J.}~\bibnamefont {Simon}}, \bibinfo {author} {\bibfnamefont {O.}~\bibnamefont {Zilberberg}},\ and\ \bibinfo {author} {\bibfnamefont {I.}~\bibnamefont {Carusotto}},\ }\bibfield  {title} {\bibinfo {title} {Topological photonics},\ }\href {https://doi.org/10.1103/RevModPhys.91.015006} {\bibfield  {journal} {\bibinfo  {journal} {Rev. Mod. Phys.}\ }\textbf {\bibinfo {volume} {91}},\ \bibinfo {pages} {015006} (\bibinfo {year} {2019})}\BibitemShut {NoStop}%
\bibitem [{\citenamefont {Chen}\ \emph {et~al.}(2025)\citenamefont {Chen}, \citenamefont {Zhang}, \citenamefont {Zou}, \citenamefont {Sun},\ and\ \citenamefont {Zhang}}]{2024arXiv240909919C}%
  \BibitemOpen
  \bibfield  {author} {\bibinfo {author} {\bibfnamefont {T.}~\bibnamefont {Chen}}, \bibinfo {author} {\bibfnamefont {W.}~\bibnamefont {Zhang}}, \bibinfo {author} {\bibfnamefont {D.}~\bibnamefont {Zou}}, \bibinfo {author} {\bibfnamefont {Y.}~\bibnamefont {Sun}},\ and\ \bibinfo {author} {\bibfnamefont {X.}~\bibnamefont {Zhang}},\ }\bibfield  {title} {\bibinfo {title} {Engineering topological states and quantum-inspired information processing using classical circuits},\ }\href {https://doi.org/https://doi.org/10.1002/qute.202400448} {\bibfield  {journal} {\bibinfo  {journal} {Advanced Quantum Technologies}\ }\textbf {\bibinfo {volume} {8}},\ \bibinfo {pages} {2400448} (\bibinfo {year} {2025})}\BibitemShut {NoStop}%
\end{thebibliography}%

\end{document}